\documentclass{jpsj3}
\usepackage{graphicx,psfrag}
\usepackage{graphicx} 
\usepackage{bm}
\usepackage{booktabs}
\usepackage{amsmath,amsthm,amssymb}
\usepackage{gt13jpsj}
\usepackage{fancybox} 
\usepackage{ascmac} 
\usepackage{framed}

\usepackage{color}
\begin{document}

\title{
Phasons and excitations in skyrmion lattice
}
\author{  Gen Tatara$^1$, Hidetoshi Fukuyama$^2$  }  
\inst{
$^1$  RIKEN Center for Emergent Matter Science (CEMS)\\  
2-1 Hirosawa, Wako, Saitama, 351-0198 Japan \\
$^2$ Research Institute for Science and Technology, Tokyo University of Science\\
2641 Yamazaki, Noda, Chiba, 278-8510, Japan
}
\date{\today}

\abst{
Excitations of two-dimensional skyrmion lattice are theoretically studied based on a collective coordinate description.
Starting from the representation of skyrmion lattice in terms of three helices, we identify the canonical coordinates describing low energy excitations 
as phasons. 
The phason excitation spectra turn out to have one gapless mode with a quadratic dispersion and one massive mode, in agreement with previous studies.
We will show that there is another collective mode governing the topological nature and the stability of skyrmion lattice and that the fluctuation of this mode leads to a screening of the topological charge of the lattice.
Experimental implications of the screening effect in  microwave absorption,  topological Hall effect and depinning threshold current in metals are discussed.
}

\maketitle

\newcommand{\geff}{\overline{g}}
\newcommand{\kva}{\kv_{a}}
\newcommand{\kvb}{\kv_{b}}
\newcommand{\kvc}{\kv_{c}}
\newcommand{\htil}{\tilde{h}}
\newcommand{\mf}{M_{\rm f}}
\newcommand{\sumtd}{\int\frac{d^2r}{a^2}}
\newcommand{\Msk}{M_{\rm h}}
\newcommand{\Mskv}{\Mv_{\rm h}}
\newcommand{\muq}{\mu_q}
\newcommand{\alphaG}{\alpha_{\rm s}}
\renewcommand{\betana}{\beta_{\rm s}}

\section{Introduction}

A magnetic skyrmion is a magnetization structure in magnetic materials with spins at the core and perimeter pointing up and down, respectively. It has a topological charge, defined in two-dimensions as
\begin{align}
n\equiv \frac{1}{4\pi M^3}\int d^2r \Mv\cdot(\nabla_x\Mv\times\nabla_y\Mv), 
\end{align}
of 1, where $\Mv$ is a vector representing the magnetization.

Clusters of such structures forming a lattice were known in thin film ferromagnets under magnetic field, although the structures were called magnetic bubbles at that time 
\cite{Malozemoff79}.
The lattice of magnetic bubbles is stabilized because of competition between a uniaxial magnetic anisotropy energy and dipolar interaction energy.
Skyrmion systems in non-centrosymmetric magnets have been intensively studied recently.
Skyrmion lattices in helimagnets were experimentally discovered by neutron scattering measurements on bulk MnSi by M\"uhlbauer et al.\cite{Muhlbauer09}.
Electron transport measurements were carried out by Neubauer  et al. and topological Hall effect due to the spin Berry's phase, which is proportional to the skyrmion number, was detected in the skyrmion phase of bulk MnSi \cite{Neubauer09}.
Real-space observation of a skyrmion lattice was carried out by Yu et al. on a thin film of
Fe$_{0.5}$Co$_{0.5}$Si using Lorentz transmission electron microscopy and the phase diagram was obtained by measuring the skyrmion density \cite{Yu10}.
The sample thickness, 20 nm, is smaller than the helix period of the system, 90 nm, and thus the skyrmion structure observed is a two-dimensional one.
It was noted there that the skyrmion lattice in this thin film appears over a wide region 
of the phase diagram, including very low temperature.
Stability of the skyrmion phase in thin films  turned out to be a common feature, as was reported in other systems like FeGe \cite{Huang12}.
A skyrmion lattice in a thin film of a chiral magnetic insulator, Cu$_2$OSeO$_3$, was observed by Lorentz TEM measurement by Seki et al.  \cite{Seki12}.
The material Cu$_2$OSeO$_3$ is multiferroic, and thus the electric control of skyrmion structures is possible \cite{White12}.
Recent developments are reviewed in Ref.  \cite{Nagaosa13}.

Jonietz et al. succeeded to induce the rotation of a skyrmion lattice in bulk MnSi by applying an electric current density of $10^6$ A/m$^2$ \cite{Jonietz10}, and 
Schulz et al. induced the translational motion at the current density of the same order \cite{Schulz12}. 
The current density is about $10^{5}$ times smaller than the typical current density needed to drive magnetic domain walls, and such low current operation is a notable feature of topological magnetic structures \cite{TKS_PR08}.
Schulz et al. detected the motion by measuring the emergent electric field generated by the moving topological object, and demonstrated that the skyrmion system is an intriguing playground for studying the emergent electromagnetism. 

The excitation modes of skyrmion lattice were observed in Cu$_2$OSeO$_3$ by microwave absorption by Onose et al. \cite{Onose12}. Excitation modes were found near 1 GHz and 1.5 GHz when the AC magnetic field is applied within and perpendicular to the skyrmion plane, respectively, and they are assigned to be the rotational and breathing modes proposed in Ref. \cite{Mochizuki12}, respectively.

A theoretical ground for stabilization of skyrmion lattices was given by M\"uhlbauer et al., based on a representation of a skyrmion lattice as a superposition of three helices \cite{Muhlbauer09}.
They discussed that the quartic term in the Ginzburg-Landau free energy, $M^4$, gives rise to a term cubic in magnetization, $BM_zM^2$, when an external magnetic field, $B$, is applied ($z$ represents the direction of the field), and have shown that this ``cubic'' term is crucial for stabilization of the three helices state.
The important role of the ``cubic`` term to stabilize the states with three density waves was pointed out originally in the case of charge-density waves (CDW) by McMillan in 1975 for transition-metal dichalcogenides where transition metals are on planar 
hexagonal lattice and  band structure with sixfold symmetry in the 
basal plane \cite{McMillan75}. 
The excitation of skyrmion lattice was studied including the  ``cubic'' term by Petrova and Tchernyshyov, who found that the two vibration modes of the lattice are mixed because of the topological nature of the skyrmion lattice, resulting in a gapped mode and a gapless mode with a quadratic dispersion\cite{Petrova11}.
Numerical studies of skyrmion lattice were carried out in Refs. \cite{Mochizuki12,Ohe13}.
The effect of a uniaxial magnetic anisotropy on stability of a skyrmion lattice (a magnetic vortex state) was studied in Refs. \cite{Bogdanov94,Wilson14}.
Current-induced dynamics of skyrmions was studied by numerical simulations \cite{IwasakiNC13,Iwasaki13,Sampaio13,Mochizuki14}.
The velocity of skyrmion lattice was deduced from a numerical simulation in Ref. \cite{IwasakiNC13}, and it was found  that the current density needed to drive a skyrmion lattice is lower than that of domain walls by a factor of about $10^{5}$. 
This feature is mainly attributed to the absence of intrinsic pinning effect for general two-dimensional topological structures \cite{TKS_PR08}.
Skyrmions may have advantage even over vortices because isolated skyrmions are repelled by sample edges at low current density \cite{IwasakiNC13,Sampaio13}, while vortices are larger and softer structures which tend to annihilate easily at the edges \cite{NSTKTM08}.
The effect of a thermal gradient was studied numerically in Ref. \cite{Mochizuki14}

Although skyrmion lattices are unique systems realized in magnets without inversion symmetry, the behaviors of excitations have similarity to those in Wigner crystals and CDW in two-dimensions under a magnetic field.
In fact, the field-induced splitting of the transverse and longitudinal phonon modes having a linear dispersion  in two-dimensional Wigner crystal was discussed in Refs. \cite{Chaplik72,Fukuyama75}.  
Under a magnetic field, one of the modes was shown to acquire a mass proportional to the magnetic field as a result of the cyclotron motion, while the other modes remain massless.
In the case of CDW, such dynamics is properly described by field 
variables (not particle coordinates) called phasons, which were first 
introduced by Lee, Rice and Anderson \cite{Lee74} for one-dimesional 
CDW resulting from Peierls transition. In fact the similar coupling of 
modes in two-dimesional CDW under magnetic field has been identified in 
terms of phasons \cite{Fukuyama78a}. Phasons are also useful to study pinning due to 
impurities and commensurability \cite{FukuyamaTakayama85}.
Regarding the pinning by random impurities, the existence of two typical cases of weak and strong pinning has been clarified for one-dimensional CDW in terms of phason by Fukuyama and Lee \cite{Fukuyama78,FukuyamaTakayama85}.
It was shown that even in the weak pinning case, a collective pinning occurs resulting in the appearance of a characteristic length scale above which the structure is disordered due to the random potentials. 
The idea was extended to the case of three-dimensions by Lee and Rice \cite{Lee79}.

Similarly, in order to describe the dynamics of magnetic structures, collective coordinates representation extracting the low energy excitations is highly useful. The dynamics of each localized spin is governed by a torque equation, and thus the dynamics of the structure may seem complicated if described in terms of the local torque acting on each spin. In contrast, the low energy behavior becomes clearer by introducing correct collective coordinates. In the case of a planar domain wall, Slonczewski pointed out that two coordinates, the wall position and the tilt angle of the wall plane, are good variable and derived the equations of motion (called the Slonczewski's equation) in the presence of an external magnetic field \cite{Slonczewski72}. 
The critical feature of the domain wall is that the two coordinates are canonically conjugate to each other. For a translational motion of the wall, therefore, tilting of the wall plane is necessary. 
This feature was shown to result, in the case driven by an electric current, in an intrinsic pinning effect, which hinders the wall motion at low current even in the absence of extrinsic pinning potentials \cite{TK04}.
It was discussed that a magnetic vortex in a film is described by the two coordinates representing the position, $X$ and $Y$, of its core in the film, and that the canonical relation between $X$ and $Y$ arises due to the topological number of the vortex \cite{Thiele73}.
The vortex is therefore free from the intrinsic pinning effect \cite{TKS_PR08}.

In this paper, we  study the dynamics of the skyrmion lattice in a two-dimensional helimagnet based on a collective coordinates description. The skyrmion lattice is described by use of three helices \cite{Muhlbauer09}, and three phase modes (phasons) representing translational modes and three modes describing the excitations out-of-the-helix planes are introduced following the study by Petrova and Tchernyshyov \cite{Petrova11}. 
The coordinates for the skyrmion lattice are field variables that depend on spatial position and the time, in contrast to the case of a domain wall and a vortex.
The field representing the phase fluctuation is called the phason.
We derive the effective Lagrangian for the collective coordinates in the slowly-varying case. 
We show that the system has two independent excitation modes corresponding to phasons along the two orthogonal directions. 
Their effective Lagrangian is shown to be equivalent to the one for a two-dimensional charged field with a mass and under a magnetic field as has been pointed out previously \cite{Chaplik72,Fukuyama75}. 
The low energy excitations are a gapless mode with quadratic dispersion and a massive mode.
The quadratic dispersion is different from the dispersion in the case of the Wigner crystal, and is due to the charge neutrality of phasons in the present case.  
The massive mode corresponds to the rotational mode of skyrmion cores.
It turns out also that there is another mode, $\varphi_0$, which is non-dynamic in the present approximation but governs the topological nature of skyrmion lattice. We show that the magnetic field applied perpendicular to the skyrmion plane suppresses the fluctuation of the $\varphi_0$ mode and this effect stabilizes the skyrmion lattice. 
When the field is weak, $\varphi_0$ fluctuation leads to a screening of the skyrmion topological number, and this screening effect affects various physical quantities such as the excitation energy, topological Hall effect and the depinning threshold current in metals.

\section{Model}
We consider a two-dimensional skyrmion lattice in the $xy$ plane based on the Ginzburg-Landau free energy for a coarse-grained magnetization in the continuum  \cite{Muhlbauer09}
\begin{align}
H &= \sumtd \lt[\frac{J}{2}(\nabla\Mv)^2
   + D\Mv\cdot(\nabla\times\Mv) \rt] +H_{M} +H_{\rm pin}+H_{\rm ST}
   \nnr
   &=\sumtd\lt[\frac{J}{2}(\nabla\Mv)^2
   - D\sum_{\mu=x,y}\sum_{\nu\lambda}\epsilon_{\mu\nu\lambda}M_\nu \nabla_\mu M_\lambda \rt] +H_{M} +H_{\rm pin}+H_{\rm ST},   \label{Hamiltonian}
\end{align}
where $\Mv$ a dimensionless vector representing the magnetization direction 
($M$ is the magnetization divided by $\frac{2\mub}{a^3}$ (for the g-factor of 2), where $a$ is the atomic spacing and $\mub$ is the Bohr magneton), $J$ is the exchange interaction energy, and $D$ is the strength of the Dzyaloshinskii-Moriya (DM) interaction.
We consider the case where the DM interaction is isotropic in the $xy$ plane.
$H_{M}$ is the free energy for the uniform magnetization including the effect of a magnetic field, $B$, applied in the $-z$ direction,
\begin{align}
H_{M} &= \sumtd \lt[\frac{a_M}{2}M^2+\frac{b_M}{4}M^4 + \frac{2\mub}{a^3}BM_z\rt],   
\end{align}
where $a_M$ and $b_M$ are parameters generally dependent on the temperature.
The effect of uniaxial magnetic anisotropy energy along the $z$-direction discussed recently \cite{Wilson14} is incorporated in the present formalism by the uniform component of magnetization ($\mf$ below) including the anisotropy.
A pinning effect due to random impurities is represented by $H_{\rm pin}$ and driving force due to electric current in metals is  represented by the spin-transfer term, $H_{\rm ST}$.
Our study without the spin-transfer term applies both to metals and insulators.

A calculation is carried out based on the Lagrangian formalism, where the Lagrangian is defined as 
\begin{align}
  L &=L_{\rm B}-H,
\end{align}
where $L_{\rm B}$ is the spin Berry's phase term, which describes the dynamics of spin systems \cite{Auerbach94}.
It is usually represented using polar coordinates for $\Mv$, $\theta, \phi$, as 
\begin{align}
  L_{\rm B}&=\sumtd \hbar M\dot{\phi}(\cos\theta-1).\label{LBdef}
\end{align}

\begin{figure}[tb]\centering
\includegraphics[width=0.45\hsize]{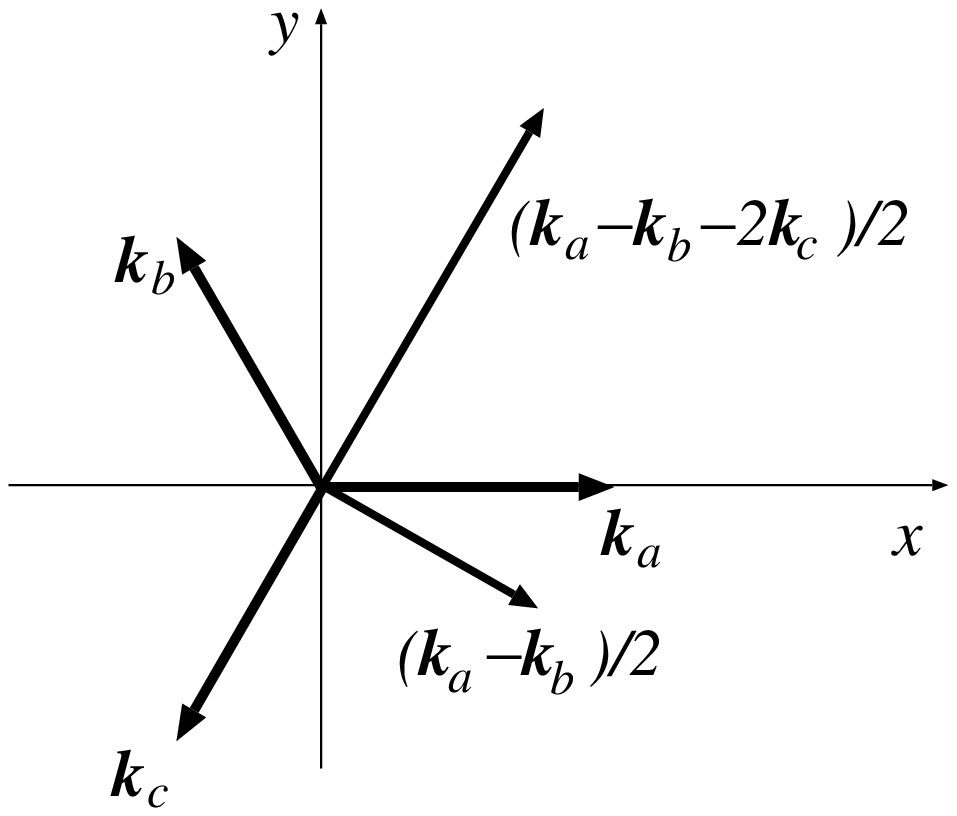}
\caption{ The wave vectors of the three helices, $\kva$, $\kvb$ and $\kvc$. \label{FIGkabc}}
\end{figure}

\begin{figure}\centering
\includegraphics[width=0.4\hsize]{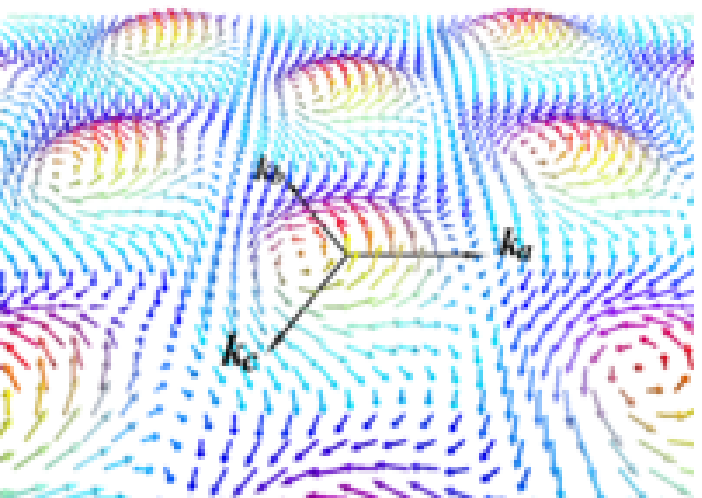}
\caption{ Plot of skyrmion structure $\Mv$ constructed as a superposition of three helices and uniform magnetization $\mf$ with $\mf/\Msk=-0.8$.  Directions of $\kva$, $\kvb$ and $\kvc$ are shown by black arrows. \label{FIGskx}}
\end{figure}

The DM interaction breaks the inversion symmetry and favors helix magnetization structures.
For instance, in the absence of external magnetic field, there is a solution of a single helix with wave vector $\kv$ with magnetization direction, $\nv$, rotating within the plane perpendicular to $\kv$. 
The wave vector may point in any direction in the $xy$-plane and its magnitude is determined by the exchange energy and DM interaction as
\begin{align}
  k &= \frac{D}{J}.
\end{align}
The chirality of the helix is determined by the DM interaction to satisfy
$(\hat{\kv}\cdot\nabla)\nv=\kv\times \nv$, where $\hat{\kv}\equiv \kv/k$, and thus a single helix solution is 
\begin{align}
  \nv_{\rm 1h} &= \hat{\zv} \cos(\kv\cdot\rv) +(\hat{\kv}\times \hat{\zv}) \sin(\kv\cdot\rv).
\end{align}
As was shown in Ref. \cite{Muhlbauer09}, the skyrmion lattice state has a lower energy than a single helix state for a finite region in the plane of the temperature and magnetic field. 
We proceed focusing on the case skyrmion lattice is realized as a ground state. 
The skyrmion lattice is represented by a superposition of three helices whose wave vectors form an equilateral triangle in the presence of a uniform magnetization component, $\mf$, induced by the applied magnetic field.
The uniform component and the interaction arising from the quartic term in $H_{M}$ are essential for the stability of three helices state
(See Ref. \cite{Muhlbauer09} and Appendix \ref{SECGL}).
The three wave vectors are chosen as (Fig. \ref{FIGkabc})
\begin{align}
\kva &=k(1,0,0) \nnr
\kvb &=k\lt(-\frac{1}{2},\frac{\sqrt{3}}{2},0\rt) \nnr
\kvc &=k\lt(-\frac{1}{2},-\frac{\sqrt{3}}{2},0\rt) ,
\end{align}
A skyrmion lattice configuration is represented by 
\begin{align}
  \Mv &= \mf\hat{\zv}+\sum_{i=a,b,c}\Mv_{i},  \label{skxconfiguration}
\end{align}
where $\Mv_{i}$ represents the magnetization vector of three helices denoted by $i=a,b,c$.
When fluctuations are neglected, we have 
$\Mv_{i} = \Msk\nv_i^{(0)}$,
where $\Msk$ is the magnitude of the helices and 
\begin{align}
  \nv_i^{(0)} &\equiv \hat{\zv} \cos(\kv_i\cdot\rv) +(\hat{\kv}_i\times \hat{\zv}) \sin(\kv_i\cdot\rv).\label{nvi0def}
\end{align}
A structure of $\Mv$ is shown in Fig. \ref{FIGskx} for $\mf/\Msk=-0.8$.
The spacing between the skyrmion cores is $a_{\rm s}\equiv\frac{4\pi}{\sqrt{3}k}=\frac{2}{\sqrt{3}}\lambda$, where $\lambda\equiv \frac{2\pi}{k}$ is the helix period. 
Since the topological charge of one skyrmion is $2\pi\frac{\hbar}{e}$, and each triangle of area $\frac{\sqrt{3}}{4}a_{\rm s}^2$ contains a half skyrmion, the average density of the effective magnetic field is $\frac{h}{e}\frac{\sqrt{3}}{2\lambda^2}$.
%

%

To describe the excitations and dynamics of the skyrmion lattice, we introduce two collective coordinates for each helix. One is $\varphi_i(\rv,t)$ representing the phase of the helices and the other is $\beta_i(\rv,t)$ representing the fluctuation out-of helix plane, i.e., the fluctuation along $\kv_i$.
The magnetization of the helices then read
\begin{align}
  \Mv_{i} 
  &= \Msk\lt( \beta_i \hat{\kv}_i +\sqrt{1-\beta_i^2} \nv_{i}\rt),
  \label{Mvi}
\end{align}
where $\nv_{i}$ are 
\begin{align}
  \nv_{i} &= \hat{\zv} \cos(\kv_i\cdot\rv+\varphi_i) +(\hat{\kv}_i\times \hat{\zv}) \sin(\kv_i\cdot\rv+\varphi_i). \label{nvidef}
\end{align}
The three variables $\varphi_i$'s represent the phasons of helical waves.

\begin{figure}[tb]\centering
\includegraphics[width=0.45\hsize]{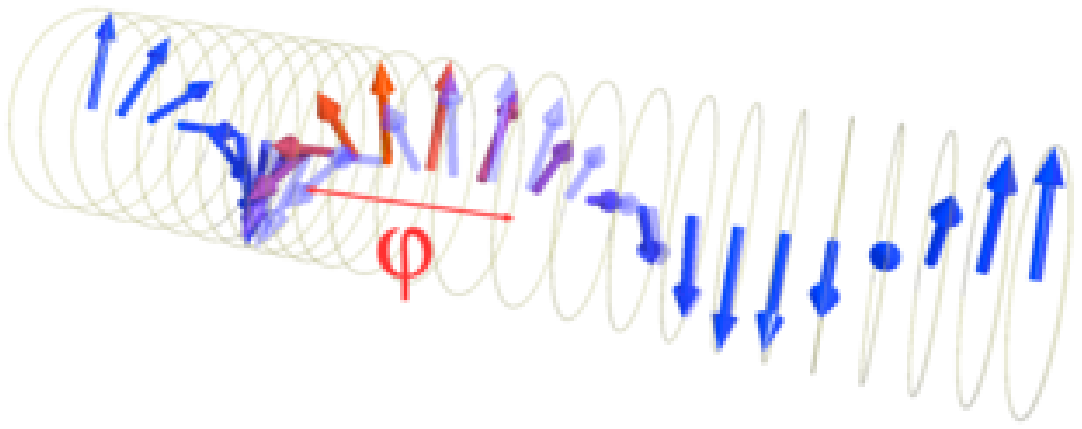}
\includegraphics[width=0.45\hsize]{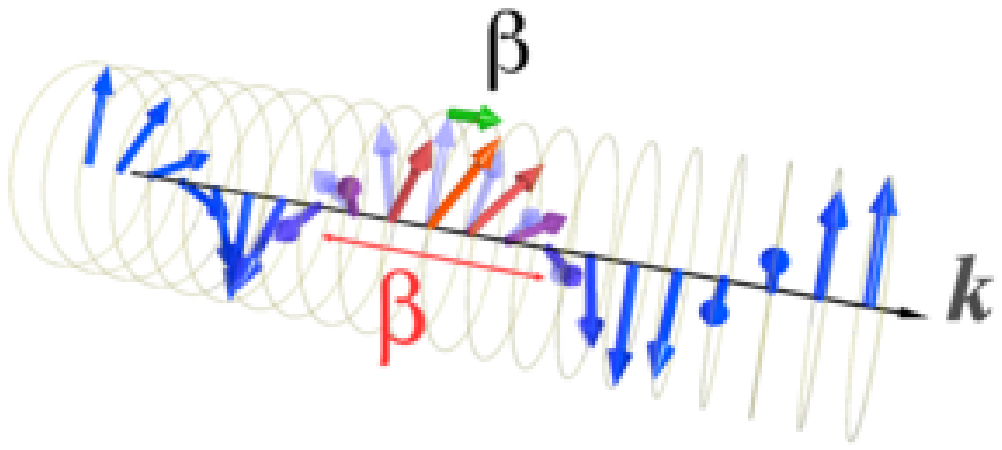}
\caption{ Excitation modes of a helix represented by $\varphi$ and $\beta$. 
Spins excited (i.e., having non-vanishing $\varphi$ or $\beta$) are shown in red. 
The mode $\varphi$ corresponds to a local modification of the helix pitch and the mode $\beta$ describes the tilt of the spin along the helix wave vector.
\label{FIGphibeta}}
\end{figure}

\section{Effective Lagrangian for collective coordinates}

We use Eq. (\ref{nvidef}) to derive the effective Lagrangian for the collective coordinates, $\varphi_i$'s and $\beta_i$'s.
We expand with respect to $\beta_i$'s to the second order, and neglect spatial variation of $\Msk$.
We assume that the fluctuations are slowly varying compared to the wave vectors of helices, and drop rapidly oscillating terms when we integrate over space.
Namely, we approximate 
\begin{align}
  \int d^2r \cos\Phi_i \cos\Phi_j &= 
  \frac{1}{2}\int d^2r\lt[\cos( (\kv_i+\kv_j)\cdot\rv+\varphi_i+\varphi_j) +  \cos( (\kv_i-\kv_j)\cdot\rv+\varphi_i-\varphi_j) \rt]\nnr
  & \simeq \frac{1}{2}\delta_{ij} \int d^2r  \cos(\varphi_i-\varphi_j) \nnr
  & = \int d^2r \sin\Phi_i \sin\Phi_j ,
\end{align}
and 
\begin{align}
 \int d^2r \cos\Phi_i \sin\Phi_j &=0  ,\label{slowvarying}
\end{align}
where $\Phi_i\equiv \kv_i\cdot\rv+\varphi_i$.
The average of the exchange interaction term then becomes
\begin{align}
  \sumtd \frac{J}{2}(\nabla\Mv)^2 
  & \simeq \sumtd \frac{J\Msk^2}{2}\lt[
  \frac{1}{2}\sum_{i\neq j} \lt(\nabla\beta_i-\nabla\beta_j\rt)^2 
  +\sum_i(1-\beta_i^2)[(k_i)^2+(\nabla\varphi_i)^2+2\kv_i\cdot\nabla\varphi_i]\rt].
\end{align}
DM term similarly is calculated as
\begin{align}
  \sumtd  D\Mv\cdot(\nabla\times\Mv)
  & \simeq   \sumtd \Msk^2 \lt[
  -3Dk+Dk\sum_i\beta_i^2-D\sum_i(\hat{\kv}_i\cdot\nabla)\varphi_i\rt].
\end{align}
The sum of the exchange and DM interactions thus reads
\begin{align}
 H_J+H_{\rm DM}
  & \simeq   \sumtd \Msk^2\lt[
  \frac{J}{2}\sum_i(k^2\beta_i^2+(\nabla\varphi_i)^2)
  +\frac{J}{4}\sum_{i\neq j} \lt(\nabla\beta_i-\nabla\beta_j\rt)^2 
  \rt].
  \end{align}
We see that the excitation mode described by $\beta_i$ has a mass proportional to $Jk^2$.

The contributions from $H_{M}$ are calculated similarly. We neglect here the contributions including massive mode, $\beta_i$.
As noted in Ref. \cite{Muhlbauer09}, the quartic term is essential for describing a skyrmion lattice.
From the expression
\begin{align}
M^4 &=  [(\mf\hat{\zv}+\Mskv)^2]^2 \nnr
&=\mf^4+\Msk^4+2\mf^2\Msk^2+4\mf^2(\Mskv\cdot\hat{\zv})^2+ 4\mf^3(\Mskv\cdot\hat{\zv})
+4\mf(\Mskv\cdot\hat{\zv})\Msk^2,
\end{align}
where $\Mskv\equiv\sum_i \Mv_i$ is the magnetization of three helices, 
we see that the last term linear in $\mf$ describes the interaction among three helices. 
The integral of the term is evaluated for slowly varying case as 
\begin{align}
\int d^2r (\Mskv\cdot\hat{\zv})\Msk^2 &= \frac{9}{4}\Msk^3 \int d^2r \cos(\varphi_a+\varphi_b+\varphi_c),
\label{cubicterm}
\end{align}
where we noted $\sum_{i=a,b,c}\Phi_i=\sum_{i}\varphi_i$ as a result of $\sum_{i}\kv_i=0$ and  
\begin{align}
\int d^2r \cos\Phi_a \cos\Phi_b\cos\Phi_c &= \frac{1}{4}\int d^2r \cos(\varphi_a+\varphi_b+\varphi_c) \nnr
&= - \int d^2r \cos\Phi_a \sin\Phi_b\sin\Phi_c.
\end{align}

\subsection{Spin Berry's phase term \label{SECsbp}}

For studying dynamics of collective coordinates, spin Berry's phase term, Eq. (\ref{LBdef}),
needs to be treated carefully, since this expression generally includes a physically irrelevant contribution arising from the fact that the term is not well-defined at $\cos\theta=-1$.
To discuss physical contribution, it is useful to rewrite the spin Berry's phase term to avoid unphysical divergence at $\cos\theta=-1$. 
To do this, we consider a change of $\Mskv$ when collective coordinates $\varphi_i$ ($i=a,b,c$) are changed from $\varphi_i^0$ to $\varphi_i$.
This change corresponds to a shift of the origin of the skyrmion lattice and does not have physical effect, since only the relative phase is meaningful if pinning is neglected.
Without losing generality, we chose $\varphi_i^0=0$.
The change of helix magnetization, $\delta\nv$, reads 
\begin{align}
\delta\nv=\sum_{i=a,b,c}\frac{\partial \nv_i}{\partial \varphi_i}\varphi_i
=\sum_{i=a,b,c}(\hat\kv_i\times\nv_i)\varphi_i. \label{deltanv}
\end{align}
By the change of phase, the spin Berry's phase term, $L_{\rm B}$, is modified by the amount \cite{Auerbach94}
\begin{align}
 \delta L_{\rm B} &= \hbar \Msk \int d^2 r \nv\cdot(\dot{\nv}\times\delta\nv).
\end{align}
This form of spin Berry's phase contains the correct physical dynamics of $\varphi_i$ and $\beta_i$.
In terms of $\beta_i$ and $\varphi_i$, $\dot{\nv}$ reads 
\begin{align}
  \dot{\nv} &= \sum_i\lt[ \dot{\varphi}_i(\hat\kv_i\times\nv_i)+\dot{\beta}_i \hat \kv_i \rt],
\end{align}
and thus by use of Eq. (\ref{deltanv}) 
\begin{align}
 \delta L_{\rm B} &= \hbar \Msk \sumtd \sum_{ijk}
\lt[
\dot{\varphi}_i \varphi_j [(\hat\kv_i\times\nv_i)\times (\hat\kv_j\times\nv_j)] 
+ \dot{\beta}_i \varphi_j [\hat\kv_i \times (\hat\kv_j\times\nv_j)] 
\rt]\cdot\nv_k.
\end{align}
The coefficient connecting $\dot{\varphi}_i$ and $\varphi_j$, 
$\int d^2r[(\hat\kv_i\cdot\nabla)\nv_i\times (\hat\kv_j\nabla)\nv_j]\cdot\nv_k$, is the topological skyrmion number for a triangular lattice unit defined by $\kv_i$ and $\kv_j$ up to a constant. 
It is easy to see that 
\begin{align}
[(\hat\kv_a\times\nv_a)\times (\hat\kv_b\times\nv_b)] \cdot\nv_c
&=
\frac{3\sqrt{3}}{8} \cos (\Phi_a+\Phi_b+\Phi_c) \nnr
&
+\frac{\sqrt{3}}{8} \lt[ \cos (\Phi_b+\Phi_c-\Phi_a)+ \cos (\Phi_a+\Phi_c-\Phi_b)- \cos (\Phi_a+\Phi_b-\Phi_c) \rt]
\nonumber\\
[\hat\kv_i \times (\hat\kv_j\times\nv_j)
] \cdot\nv_j
&= -(\kv_i\cdot\kv_j)= -\frac{3}{2}\delta_{i,j}+\frac{1}{2},
\end{align}
and we thus obtain, dropping oscillating contributions, 
\begin{align}
\int d^2r [(\hat\kv_i\times\nv_i)\times (\hat\kv_j\times\nv_j)] \cdot\nv_k
&=
\int d^2r \frac{3\sqrt{3}}{8} \epsilon_{ijk} \cos (\varphi_a+\varphi_b+\varphi_c) ,
\label{SBPvalue}
\end{align}
where $i,j,k$ runs over $a,b,c$.
The result of spin Berry's phase term is thus 
\begin{align}
 \delta L_{\rm B} &= \hbar g\Msk \sumtd
\lt[
(\dot{\varphi}_a \varphi_b- \varphi_a \dot \varphi_b)
+(\dot{\varphi}_b \varphi_c- \varphi_b \dot \varphi_c)
+(\dot{\varphi}_c \varphi_a- \varphi_c \dot \varphi_a)\rt]
\cos (\varphi_a+\varphi_b+\varphi_c) 
\nnr
&+ 
\hbar \Msk \sumtd 
\lt[
\dot{\beta}_a \lt(\varphi_a- \frac{1}{2}\lt(\varphi_b + \varphi_c\rt)\rt)
+
\dot{\beta}_b \lt(\varphi_b- \frac{1}{2}\lt(\varphi_c + \varphi_a\rt)\rt)
+
\dot{\beta}_c \lt(\varphi_c- \frac{1}{2}\lt(\varphi_a + \varphi_b\rt)\rt)
\rt],
\end{align}
where 
\begin{align}
  g \equiv \frac{3\sqrt{3}}{8},
\end{align}
represents the topological number density for $\varphi_i$'s.

We note that if we estimate the spin Berry's phase term based on the familiar expression (\ref{LBdef}), we obtain 
\begin{align}
  \int d^2 r \cos\theta \dot{\phi} 
  &= \int d^2 r \frac{M_z}{M_x^2+M_y^2}(\Mv\times \dot{\Mv})_z \nnr
  &= - \sum_{i}\dot\varphi_i\lt(\frac{3}{2}\beta_i-\frac{1}{2}\sum_j\beta_j\rt).
\end{align}
This result is does not have important terms connecting $\dot\varphi$ and $\varphi$, and is not correct.
This results from a straightforward treatment of $ \frac{1}{M_x^2+M_y^2}$, which diverges at $M_x=M_y=0$.
The same goes for the spin-transfer torque term.

\subsection{Lagrangian}

The total Lagrangian for the collective coordinates (in the absence of current and pinning) thus reads
\begin{align}
  L &=   \Msk^2 \sumtd \lt[ \frac{\hbar g}{\Msk} \lt[\dot{\varphi}_a(\varphi_b-\varphi_c)+\dot{\varphi}_b(\varphi_c-\varphi_a)+\dot{\varphi}_c(\varphi_a-\varphi_b)\rt]\cos(\varphi_a+\varphi_b+\varphi_c) \rt. \nnr
& 
    -\frac{\hbar}{\Msk}\lt( \dot\varphi_a\lt(\beta_a-\frac{1}{2}(\beta_b+\beta_c) \rt) + \dot\varphi_b\lt(\beta_b-\frac{1}{2}(\beta_c+\beta_a) \rt)
    + \dot\varphi_c\lt(\beta_c-\frac{1}{2}(\beta_a+\beta_b) \rt) \rt) 
    \nnr
    &
   \lt.
   -\frac{J}{2}\sum_i(k^2\beta_i^2+(\nabla\varphi_i)^2)
  -\frac{J}{4}\sum_{i\neq j} \lt(\nabla\beta_i-\nabla\beta_j\rt)^2  
 + h \cos(\varphi_a+\varphi_b+\varphi_c) \rt] , \label{Lfree}
\end{align}
where $h\equiv -\frac{9}{4}b\mf\Msk(>0)$.
The above expression was derived assuming slowly-varyingness. 
As for $\varphi_i$'s, we have not carried out a perturbative expansion to keep the periodic nature of the variable. 
In fact, $\varphi_i$'s form zero modes (in the absence of pinning) and they are not necessarily small in  amplitude.

The Lagrangian, (\ref{Lfree}), is the one in the representation using canonical coordinates and momenta.
The time-derivative term of the Lagrangian is a product of canonical momentum, $p$, and time-derivative of canonical coordinate, $\dot{q}$.
We thus see that the canonical momentum of $\varphi_a$ is a composite field, 
$g(\varphi_b-\varphi_c)\cos(\varphi_a+\varphi_b+\varphi_c)-\beta_a+\frac{1}{2}(\beta_b+\beta_c)$.
The skyrmion lattice dynamics appears thus to be complicated if represented in terms of $\varphi_i$ and $\beta_i$. 
This means that these variables themselves are not good variables to describe low energy dynamics. 
Our next task is to find good variables, which we shall carry out in the next section.

Before proceeding further, let us look into the Lagrangian more closely.
When $h$ is large, $\varphi_i$'s are constrained to satisfy $\cos(\varphi_a+\varphi_b+\varphi_c)\simeq1$, and the canonical relations are determined by the first term of Eq. (\ref{Lfree}). 
(The second term connecting $\dot{\varphi}_i$ to $\beta_i$ turns out to be irrelevant in the low energy behavior as we shall see later.)
For instance, the canonical momentum of  ${\varphi}_a$, a shift of spiral phase in the $x$ direction,  is $(\varphi_b-\varphi_c)$, a shift of phase in the $y$ direction. 
The motion of skyrmion lattice in the $x$ direction and $y$ direction are thus coupled in the same manner as a motion of charged particles in the presence of a magnetic field.
This is due to the fact that the topological number (spin Berry's phase) of skyrmion acts as an effective magnetic field.
This feature of topological magnetic structures was noted in the case of magnetic vortices by Thiele \cite{Thiele73}.
In contrast, when $h$ is small, $\varphi_i$'s may fluctuate independently, resulting in $\cos(\varphi_a+\varphi_b+\varphi_c)\sim0$ and the disappearance of the kinetic term proportional to $g$. 
The canonical momentum of $\varphi_a$ then reduces to $\lt(\beta_a-\frac{1}{2}(\beta_b+\beta_c) \rt)$ and 
the dynamics reduces to that of a phonon with a linear dispersion as we will show below.
We therefore see that the parameter $h$ is essential in determining the skyrmion dynamics.

\section{Excitations}

The Lagrangian we obtained, (\ref{Lfree}), seems complicated as it is.
We will demonstrate in this section that it becomes much simpler if we choose good dynamic variables.
Finding good variables can be realized by looking into the form of the kinetic term. 
In the present case, from the terms proportional to $g$, we expect that differences such as $\varphi_a-\varphi_b$ would be a good variable. 
After some algebra, it is easy to see that another good variable is 
 $(\varphi_a-\varphi_c)+(\varphi_b-\varphi_c)$. 
Let us thus define 
\begin{align}
\varphi_+ &\equiv \frac{1}{2}(\varphi_a+\varphi_b)-\varphi_c  \nnr
\varphi_- &\equiv \frac{1}{2}(\varphi_a-\varphi_b).
\end{align}
Note that both variables are written in terms of phase differences.
The Berry's phase term proportional to $g$ then is simplified as 
\begin{align}
\dot{\varphi}_a(\varphi_b-\varphi_c)+ \dot{\varphi}_b(\varphi_c-\varphi_a) + \dot{\varphi}_c(\varphi_a-\varphi_b) 
&= 2( \varphi_+\dot{\varphi}_- - \varphi_- \dot{\varphi}_+).
\end{align}
Similarly, the other Berry's phase contribution connecting $\dot{\varphi}_i$ and $\beta_i$ is simplified by introducing 
\begin{align}
\beta_+ &\equiv \frac{1}{2}(\beta_a+\beta_b)-\beta_c  \nnr
\beta_- &\equiv \frac{1}{2}(\beta_a-\beta_b),
\end{align}
as 
\begin{align}
\dot\varphi_a\lt(\beta_a-\frac{1}{2}(\beta_b+\beta_c)\rt) +
\dot\varphi_b\lt(\beta_b-\frac{1}{2}(\beta_c+\beta_a)\rt) +
\dot\varphi_c\lt(\beta_c-\frac{1}{2}(\beta_a+\beta_b)\rt) 
&= \beta_+ \dot{\varphi}_+ + 3 \beta_- \dot{\varphi}_-.
\end{align}
Here we notice that, although we have originally introduced six variables, $\varphi_i$'s and $\beta_i$'s, there are only four independent dynamic variables, $\varphi_\pm$ and $\beta_{\pm}$, since the kinetic terms are expressed using only these four variables. Since two of the four are the canonical momenta, we now see that there are only two excitation modes in the present effective Lagrangian.
This fact is natural since the excitations corresponding to a sliding of helix $\varphi_a$ necessarily induces the motion of the other two helices, since the the direction of the sliding of the three helices are not orthogonal.
In other words, there are only two independent spin fluctuation propagation directions in two dimensions.

\begin{figure}\centering
\includegraphics[width=0.45\hsize]{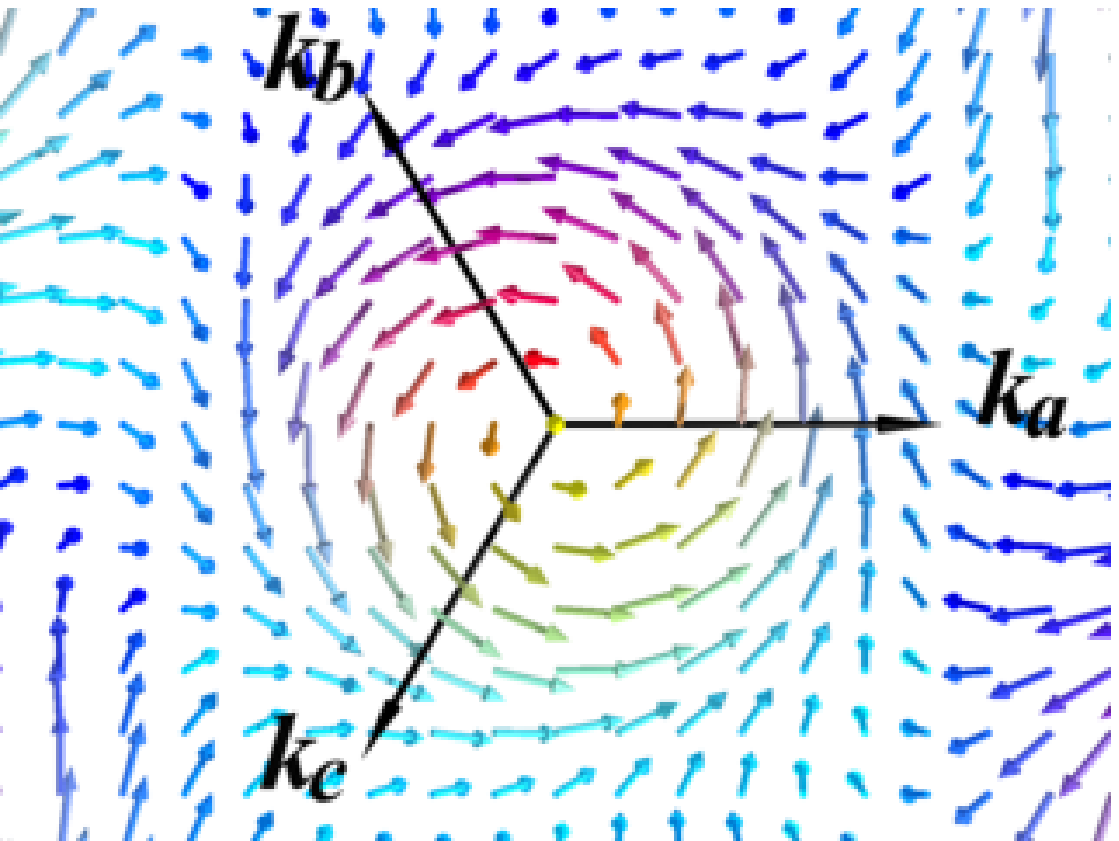}
\includegraphics[width=0.45\hsize]{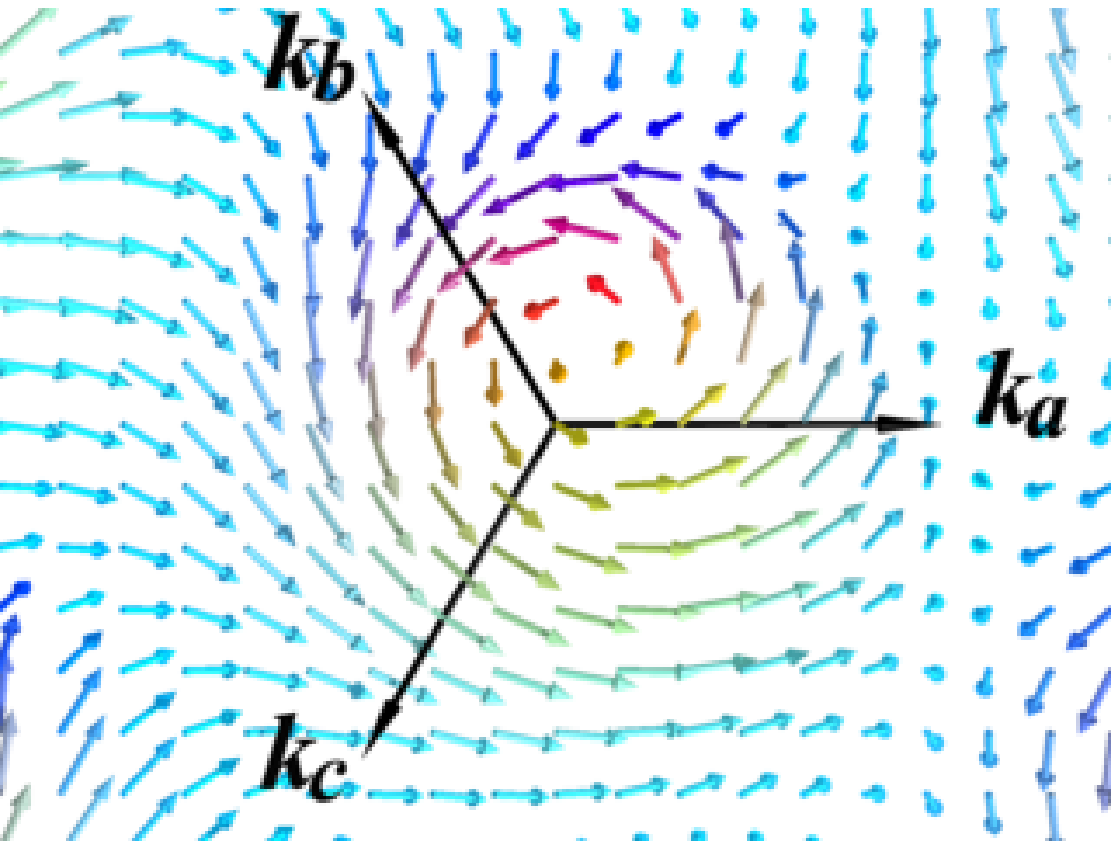}
\caption{ Skyrmion lattice structures with $\beta_-=0$ (left) and $\beta_-=-0.5$ (right).
It is seen that $\beta$ excitation corresponds to a shift of the core, i.e., a rotational mode. \label{FIGbeta}}
\end{figure}

We have succeeded in finding the correct dynamic variables in the Lagrangian. 
There are, however, two other variables in the Lagrangian. 
Although not dynamic, these variables may have essential effects on the dynamics of $\varphi_\pm$ and $\beta_\pm$.
Introducing two variables as 
\begin{align}
\varphi_0 & \equiv \varphi_a  + \varphi_b + \varphi_c  \nnr
\beta_0   & \equiv \beta_a + \beta_b +\beta_c ,
\end{align}
we see that the Lagrangian is simplified. 
For instance, 
\begin{align}
\sum_i \beta_i^2 &= \frac{2}{3}(\beta_+^2+3\beta_-^2)+\frac{1}{3}\beta_0^2 \nnr
\sum_{\langle i,j\rangle}(\nabla\beta_i-\nabla\beta_j)^2 
 &= 2[(\nabla\beta_+)^2+3(\nabla\beta_-)^2 ].
\end{align}
The Lagrangian thus reads
\begin{align}
L &= \Msk^2 \sumtd \lt[\frac{\hbar}{\Msk}\lt[ 2 g\cos\varphi_0( \varphi_+\dot{\varphi}_- - \varphi_- \dot{\varphi}_+)
   -( \beta_+ \dot{\varphi}_+ + 3 \beta_- \dot{\varphi}_- ) \rt] \rt.\nnr
 & -\frac{J}{3} \lt( (\nabla\varphi_+)^2+3(\nabla\varphi_-)^2 +\frac{1}{2}(\nabla\varphi_0)^2 \rt)  \nnr
 & \lt. 
 - \frac{J}{3}k^2 \lt(  \beta_+^2+3\beta_-^2+\frac{1}{2}\beta_0^2 \rt)
 - \frac{J}{2} [(\nabla\beta_+)^2+3(\nabla\beta_-)^2]
   +h\cos\varphi_0 \rt]. \label{Lphibeta}
\end{align}
We see that the Lagrangian is now expressed in terms of two pairs of two canonical variables, $\varphi_\pm$ and $\beta_\pm$, and two modes $\varphi_0$ and $\beta_0$.
The meaning of $\varphi_\pm$ are as follows.
The mode $\varphi_-$ describes the phase fluctuation (phason) along  $\kv_-\equiv\kva-\kvb=\frac{\sqrt{3}}{2}\lt(\sqrt{3},-1,0\rt)$, while the propagation direction of 
$\varphi_+$ is along $\kv_+\equiv\kva+\kvb-2\kvc=\frac{3}{2}\lt(1,\sqrt{3},0\rt)$ (Fig. \ref{FIGkabc}).
The two modes thus correspond to skyrmion propagation in the two orthogonal directions. 
A different factor of three in the Lagrangian is due to the difference of the magnitude of the  wavelengths, $k_+$ and $k_-$.
The mode represented by $\beta_\pm$ describes the deformation of skyrmion lattice into the configuration more like a single helix in the directions $\kv_\pm$, respectively (Fig. \ref{FIGbeta}).
The mode $\varphi_0$ affects the phason dynamics via the kinetic term (the first term of Eq. (\ref{Lphibeta})). It is not a dynamic variable in the standard sense, since the Lagrangian does not have terms containing $\dot{\varphi}_0$. 
(We may rewrite the kinetic term by use of partial integration with respect to time and obtain higher-order kinetic terms like 
$\dot{\varphi}_0\varphi_+\varphi_- \sin\varphi_0 $, but non-linear canonical momenta obtained from such kinetic terms are neglected in standard treatments.)
The mode $\beta_0$ is decoupled from other modes and is irrelevant. 

From the kinetic term of (\ref{Lphibeta}), we see that 
phason $\varphi_+$ has two canonical momenta, $\varphi_-$ and $\beta_+$, and canonical momenta for $\varphi_-$ are $\varphi_+$ and $\beta_-$.
The first canonical relation,  between $\varphi_+$ and $\varphi_-$, is induced by a topological effect, and indicates that for the motion of the phason mode $\varphi_+$, excitation of $\varphi_-$ mode is necessary. 
This behavior has been known in topological magnetic structures such as magnetic vortices \cite{Thiele73,SNTKO06}. In fact, the center of mass of a vortex in two dimensions, $X$ and $Y$ are canonically conjugate to each other and thus when a force along $x$-axis is applied, motion in the $y$-direction is induced.
The system is thus similar to that of charged particle under a magnetic field. The effective magnetic field is generated by topological number of the magnetic structure. 
The other canonical momentum for $\varphi_\pm$ is $\beta_\pm$.
This feature is similar to the case of magnetic domain wall.
In fact, for a domain wall to slide, a tilting of the wall plane is necessary  \cite{Chikazumi97,TK04,TKS_PR08}.
This is because the translational motion of a wall is induced when a torque rotating the magnetization within the wall plane is applied, and this torque needs to be generated if the magnetization component  to the wall plane is induced. 
It is a unique feature of two-dimensional skyrmion lattice, described by two phason fields, that it has both properties of domain walls and vortices.
As we saw from the derivation, the two canonical relations arise from the same spin Berry's phase term in the Lagrangian. Spin Berry's phase term in the Lagrangian is equivalent to imposing the commutation relation of the spin operators, and  
both relations are therefore different manifestations of the spin commutation relation.

\subsection{Dispersion for the small amplitude case}
Dispersion of the excitation is now easy to calculate.
Let us first consider the dispersion in the case of small amplitude of $\varphi_i$'s, expanding $\cos\varphi_0$ to the second order in $\varphi_0$.
The time-integral of the Lagrangian, i.e., the action, in the Fourier representation is
\begin{align}
   \int dt L =   \int \frac{d\omega}{2\pi} \sum_{\qv} 
  {\bm \varphi}_{-\qv,-\omega}^{\rm t} {\cal L_\varphi} {\bm \varphi}_{\qv,\omega},
  \label{Lphibeta2}
\end{align} 
where the basis is chosen as (${\rm t}$ stands for transpose) 
\begin{align}
  {\bm \varphi}_{\qv,\omega}\equiv ( \varphi_+(\qv,\omega),\varphi_-(\qv,\omega),\beta_+(\qv,\omega),\beta_-(\qv,\omega),\varphi_0(\qv,\omega),\beta_0(\qv,\omega))^{\rm t},
\end{align}
and the $6\times6$ matrix is
\begin{align}
  {\cal L_\varphi} \equiv
   \Msk ^2 \lt(  
 \begin{array}{cccccc}
 \frac{-J}{3}q^2 & -  \frac{ 2ig }{\Msk}\hbar  \omega & - \frac{i}{2\Msk}\hbar  \omega   & 0 & 0 & 0 \\
   \frac{ 2ig}{\Msk} \hbar  \omega & -J q^2 & 0 & -\frac{3i}{2\Msk}\hbar  \omega   & 0 & 0 \\
 \frac{i}{2\Msk}\hbar  \omega  & 0 & \frac{-J}{3}\left(k^2+\frac{3}{2}q^2\right) & 0 & 0  & 0 \\
 0 & \frac{3i}{2\Msk}\hbar  \omega   & 0 & -J \left(k^2+\frac{3}{2}q^2\right) & 0 & 0 \\
 0  & 0 & 0 & 0 & \frac{-J}{6}q^2-\frac{h}{2} & 0 \\
 0 & 0 & 0 & 0 & 0 & \frac{-J}{6}k^2 \\
\end{array}
\right).  \label{Lphibeta2matrix}
\end{align}
In the matrix representation, it is clear that $\beta_0$ is orthogonal to other modes and is irrelevant.
The mode $\varphi_0$ is also decoupled in the small amplitude case, but it has an important role of renormalizing the topological number, $g$, and $h$ via the factor of $\cos\varphi_0$.
We shall study the effect of the renormalization in Sect. \ref{SECrenorm} and focus here on the small amplitude case.
The determinant of the matrix is (dropping an irrelevant constant)
\begin{align}
(Jq^2+3h) [\omega^2  - (\omega_q^+)^2 ][\omega^2  - (\omega_q^-)^2 ], \label{determinant_1}
\end{align}
where 
\begin{align}
\hbar\omega_q^\pm & \equiv
\frac{4\sqrt{2}}{3\sqrt{3}}gJ\Msk k^2 \muq^{\frac{1}{2}}
 \left(\muq+\frac{3q^2}{8k^2g^2}\right)^{\frac{1}{2}}
 \lt[ 1
  \pm \frac{ \muq^{\frac{1}{2}} \lt( \muq+\frac{3q^2}{4k^2g^2} \rt)^{\frac{1}{2}}} { \muq+\frac{3q^2}{8k^2g^2} } 
  \rt]^{\frac{1}{2}} ,
  \label{dispersion}
\end{align} 
where 
$\muq\equiv \left(1+\frac{3q^2}{2k^2}\right)$.
\begin{figure}\centering
\includegraphics[width=0.45\hsize]{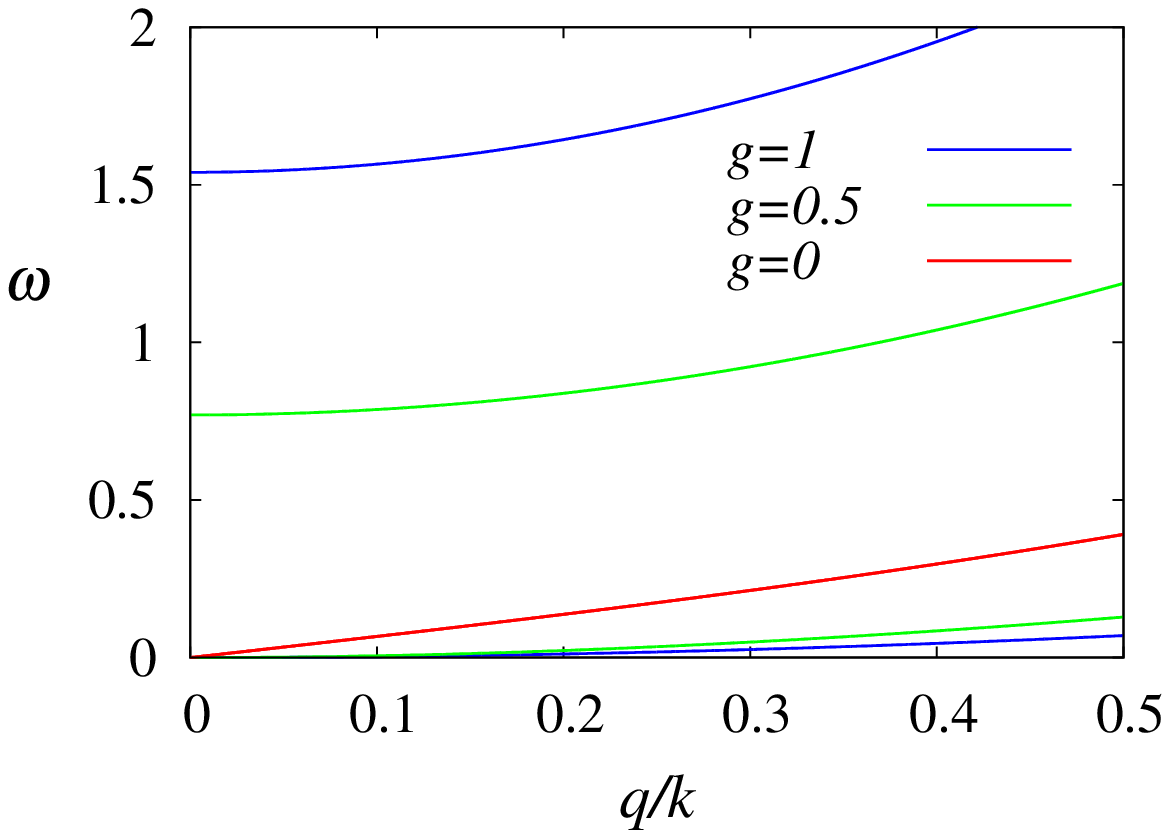}
\includegraphics[width=0.45\hsize]{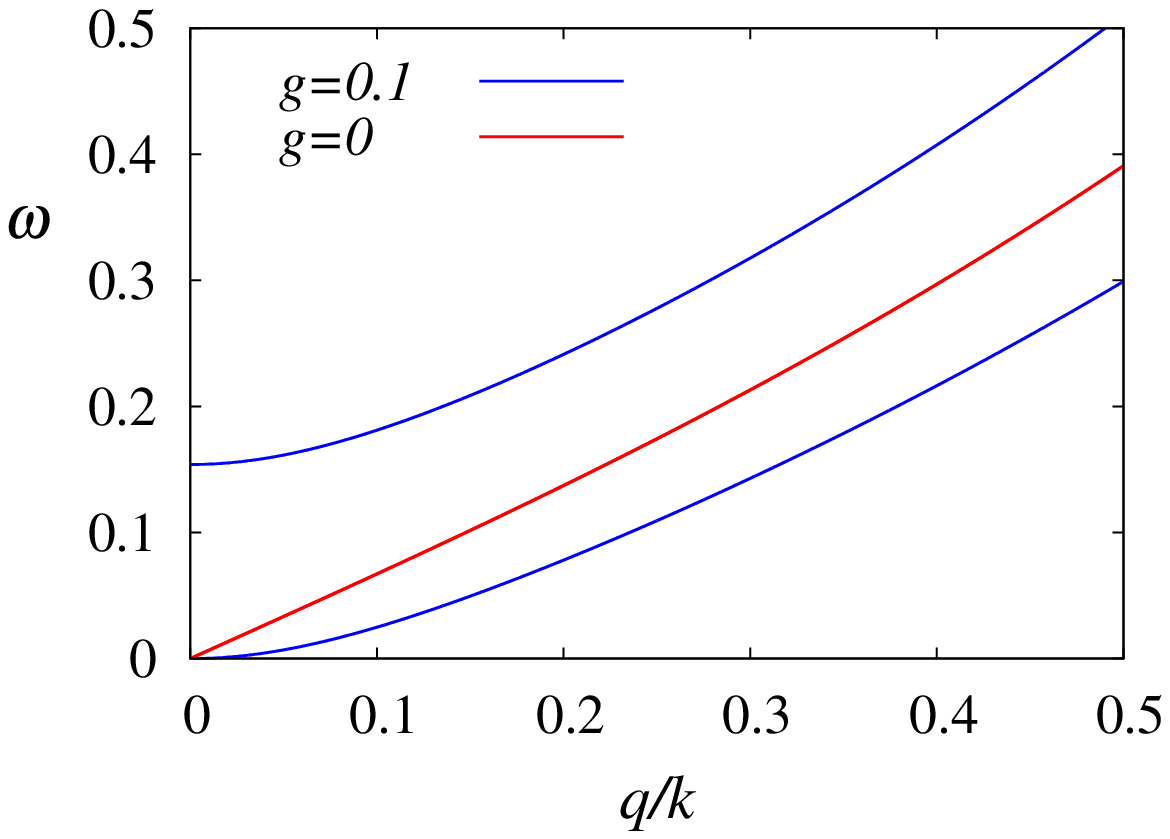}
\caption{ Left: Plot of dispersion of excitations, $\omega_q$ in unit of $J\Msk k^2/\hbar$ as function of $\frac{q}{k}$ for $g=1, 0.5,0$.  
Right:  Plot of dispersion of excitations for small $g$ ($g=0.1,0$). The crossover to linear dispersion occurs at $\frac{q}{k}\sim g$. 
\label{FIGdispersion}}
\end{figure}
If $g=0$, Eq. (\ref{dispersion}) leads to 
\begin{align}
\hbar \omega_q = \frac{2}{3}J\Msk kq\sqrt{1+\frac{3q^2}{2k^2}} \equiv \omega_q^{\rm L}.
\label{omegaL}
\end{align}
The degenerate linear dispersion indicates the existence of two vibration modes of spins in two-dimensions similar to the phonon system.
Once $g\neq0$, there is a gapless mode with dispersion $\omega\propto q^2$ and a massive mode as $q\ra0$  (Fig. \ref{FIGdispersion}) as seen from Eq. (\ref{dispersion}) expanded to the order of $q^2$, 
\begin{align}
\hbar\omega_q^\mp&= \lt\{ 
        \begin{array}{c}
           \frac{J\Msk}{2\sqrt{3}g}q^2 \\
           \frac{8gJ\Msk k^2}{3\sqrt{3}} +\frac{J\Msk (1+8g^2)}{2\sqrt{3}g}q^2.
         \end{array} \rt. \label{dispersion_q2}
\end{align}
The singular behavior of the prefactors of $q^2$ terms at $g\ra0$ is a signature of crossover from a quadratic dispersion to a linear one. 
Effects of finite $g$ are characterized by the parameter $q/(kg)$.
When $g$ is finite, $\varphi_+$ and $\varphi_-$ starts to form canonical conjugates due to the topological nature of skyrmions. In other words, the effective magnetic field arising from the spin Berry's phase induces the Lorentz force on the $\varphi_\pm$ modes and mixes the two modes, resulting in the softening of one of the linear modes and in the formation of a massive mode with a gap proportional to $g$.

The mechanism of softening in the presence of magnetic field is the same as the one in the case of the Wigner crystal discussed in Refs. \cite{Chaplik72,Fukuyama75,Fukuyama78a}, but the quadratic dispersion is distinct from the case of the Wigner crystal having the  $q^{\frac{3}{2}}$ dispersion, because of the neutral charge of the phasons in the present case.
Another unique feature of the skyrmion lattice is that the effect of $g$ is non-perturvative in the sense that the high energy behavior at $q$ larger than the crossover value ($\sim kg$) is also affected by $g$; the linear dispersions at $g\ll1$ and $q\gtrsim kg$,
\begin{align}
  \hbar\omega_q^\mp&= 
           \frac{2}{3}{J\Msk}kq\lt(1\pm \frac{2}{\sqrt{3}}\frac{kg}{q} +O\lt(\lt(\frac{kg}{q}\rt)\rt)^2\rt) ,
\end{align}
do not merge to a single line for small but finite $g$ as seen in Fig. \ref{FIGdispersion}, in contrast to the case of the Wigner crystal.

\section{Renormalization of the topological term \label{SECrenorm}}

\begin{figure}\centering
\includegraphics[width=0.4\hsize]{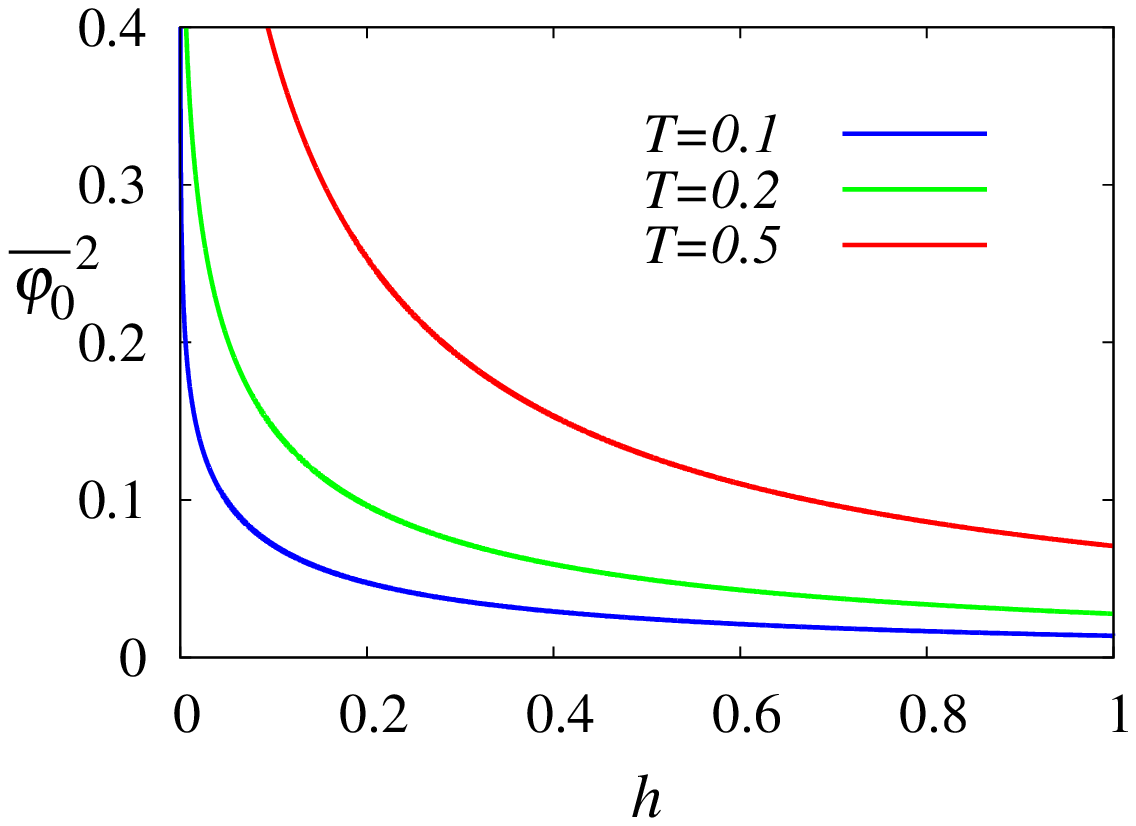}
\includegraphics[width=0.4\hsize]{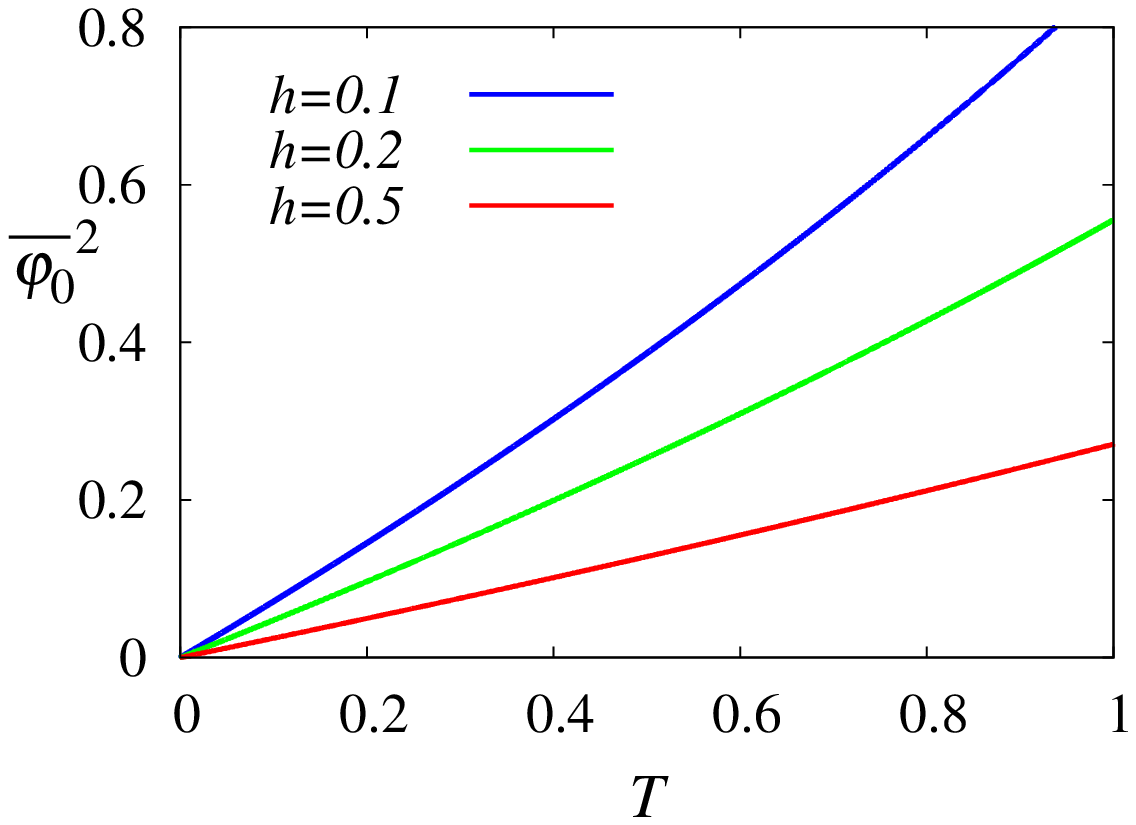}
\caption{ Left: Plot of $\overline{\varphi_0^2}$ as function of scaled magnetic field, $h$, for $T=0.1$, $T=0.2$ and $T=0.5$.
Right: Plot of $\overline{\varphi_0^2}$ as function of the temperature, $T$, for $h=0.1$, $h=0.2$ and $h=0.5$.
Temperature is in unit of $J/\kb$ and $h$ is in unit of $Jk^2$.
\label{FIGphi_hT}}
\end{figure}

\begin{figure}\centering
\includegraphics[width=0.4\hsize]{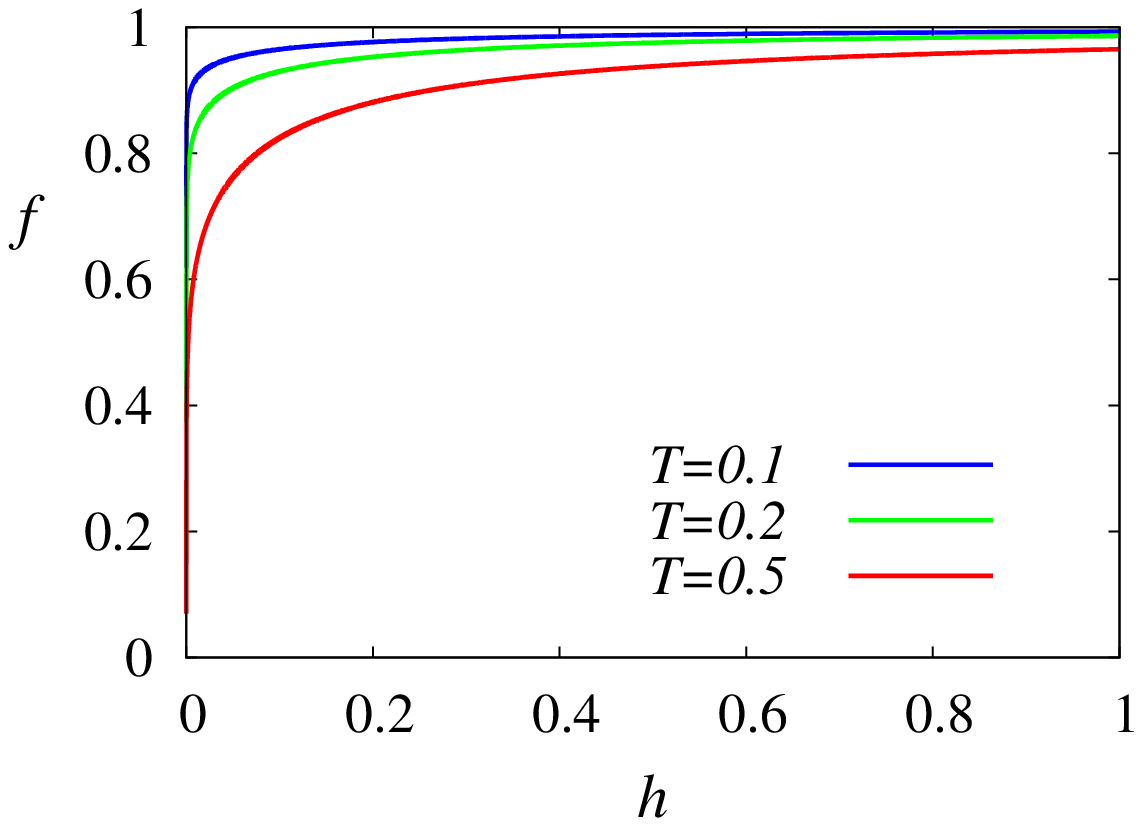}
\includegraphics[width=0.4\hsize]{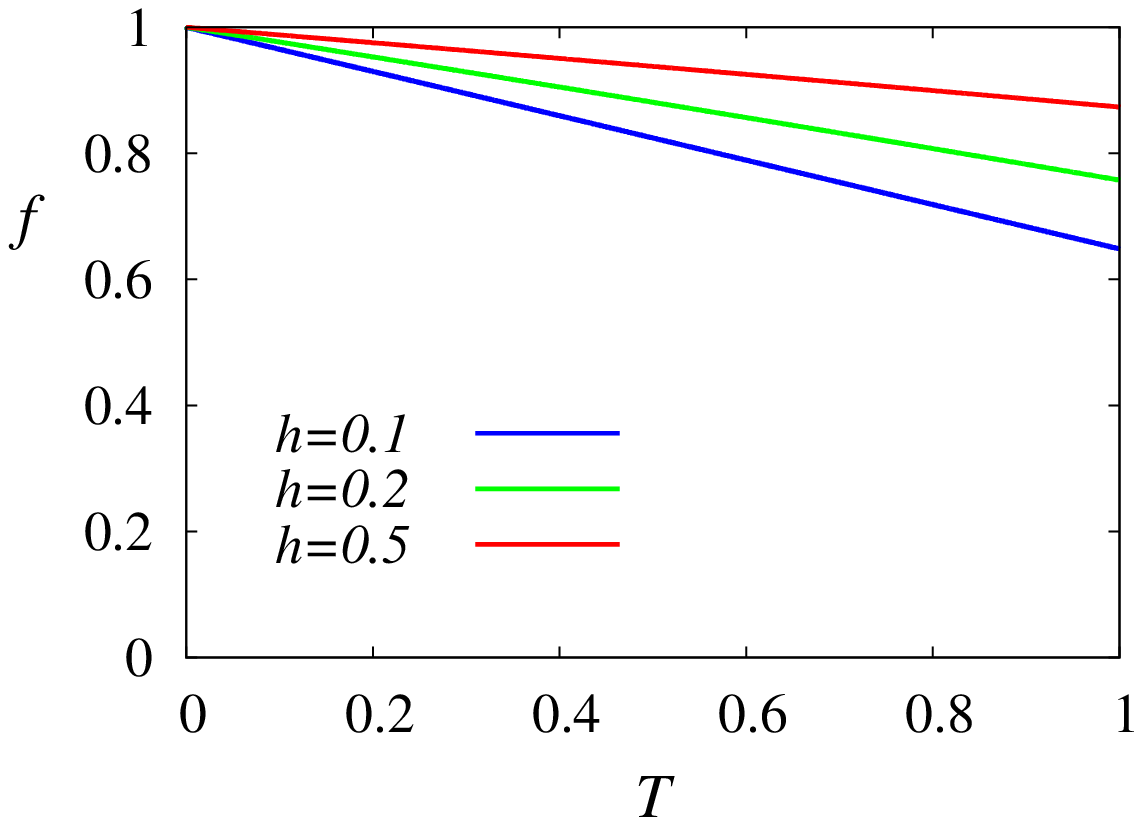}
\caption{ Left: Plot of $f=e^{-\frac{\overline{\varphi_0^2}}{2}}=\geff/g$ as function of $h$ for $T=0.1$, $T=0.2$ and $T=0.5$.
Right: Plot of $f$ as function of $T$ for $h=0.1$, $h=0.2$ and $h=0.5$. \label{FIGf_hT}}
\end{figure}

As seen in the Lagrangian, (\ref{Lphibeta}), the $\varphi_0$ mode has an important effect on the dynamics of $\varphi_\pm$ by renormalizing the topological ($g$-)term. 
To take account of this renormalization effect, we treat $\varphi_0$ beyond the second order expansion scheme by use of self-consistent harmonic approximation \cite{Dashen74,Nakano80}.
This approximation takes account of the renormalization effect by use of the expectation value of $\varphi_0^2$, $\overline{\varphi_0^2}$ and take account of dynamic fluctuation as small variable.
We thus rewrite $\cos\varphi_0$ as
\begin{align}
\cos({\varphi_0}) &= 1-\frac{1}{2}\varphi_0^2 +\cdots \nnr
  &= 1-\frac{1}{2} \overline{\varphi_0^2}
    -\frac{1}{2}(\varphi_0^2 - \overline{\varphi_0^2} )  +\cdots \nnr
     &\simeq f
     \lt(1-\frac{1}{2}(\varphi_0^2 - \overline{\varphi_0^2})
     \rt),
\end{align}
where 
\begin{align}
f\equiv e^{-\frac{1}{2} \overline{\varphi_0^2} }.
\end{align}
The topological term then becomes 
$2\hbar fg( \varphi_+\dot{\varphi}_- - \varphi_- \dot{\varphi}_+)$, and the Lagrangian for $\varphi_0$ is
\begin{align}
  L_{{\varphi_0}} 
  & \equiv  \sumtd \frac{1}{2}\lt[ -\frac{J}{3}(\nabla \varphi_0)^2   
  - hf \varphi_0^2
   \rt]. \label{Lvarphi0}
\end{align}
The average $\overline{\varphi_0^2}$ at the temperature $T$ is calculated as (see Appendix \ref{SECphiint}) 
\begin{align}
\overline{{\varphi_0}^2} 
 &= 6\kb T \sumqv \frac{1}{{J}q^2+3hf},
 \label{averagephi2}
\end{align}
where $\kb$ is the Boltzmann constant.
We carry out the integration over $\qv$ in two-dimensions, introducing a large wavelength cutoff of $k$, as 
\begin{align}
\sum_{\qv}\frac{1}{Jq^2+3hf }
&= \frac{1}{4\pi}\int_0^{k^2} \frac{dq^2}{{J}q^2+3hf} \nnr
&= \frac{1}{4\pi J}\ln\lt(1+\frac{Jk^2}{3hf}\rt).
\end{align}
The self-consistency condition, Eq. (\ref{averagephi2}), thus reads 
\begin{align}
\overline{{\varphi_0}^2}
 &=\frac{3\kb T}{2\pi J}\ln\lt(1+\frac{Jk^2}{3h e^{-\frac{\overline{{\varphi_0}^2}}{2}}}\rt)
 \label{sceq1}
\end{align}
or
\begin{align}
f
 &=\lt(1+\frac{Jk^2}{3hf}\rt)^{ -\frac{3\kb T}{4\pi J} }.
 \label{schaf}
\end{align}
Near $T=0$, 
\begin{align}
f
 &=1- \frac{3\kb T}{4\pi J}\ln\lt(1+\frac{Jk^2}{3h}\rt)+O(T^2).
\end{align}
The solutions of $\overline{{\varphi_0}^2}$ and $f$ are plotted in Figs. \ref{FIGphi_hT} and \ref{FIGf_hT}.

The Lagrangian after self-consistent harmonic approximation reads
\begin{align}
L &= \Msk^2 \sumtd \lt[\frac{\hbar}{\Msk}\lt[ 2 \geff ( \varphi_+\dot{\varphi}_- - \varphi_- \dot{\varphi}_+)
   -( \beta_+ \dot{\varphi}_+ + 3 \beta_- \dot{\varphi}_- ) \rt] \rt.\nnr
 & \lt.
 -\frac{J}{3} \lt( (\nabla\varphi_+)^2+3(\nabla\varphi_-)^2 \rt)  
 - \frac{J}{3} \lt( k^2 ( \beta_+^2+3\beta_-^2) \rt)
           -\frac{J}{2} [(\nabla\beta_+)^2+3(\nabla\beta_-)^2]
   \rt], \label{Lphibetascha}
\end{align}
where $\geff\equiv fg$ is the effective topological amplitude and $f$ is a solution of Eq. (\ref{schaf}).
The excitation energy including the renormalization effect is obtained by replacing $g$ in Eqs. (\ref{dispersion})(\ref{dispersion_q2}) by $\geff$.

\subsection{Cubic term and stability of skyrmion lattice}
In the present scheme based on a description using three helices, the ``cubic'' term (Eq. (\ref{cubicterm}) and the term proportional to $h$ in Eq. (\ref{Lphibeta})) is essential for stabilization of the skyrmion lattice. 
In fact, if $h=0$, the amplitude of $\varphi_0$ diverges in two-dimensions according to Eq. (\ref{averagephi2}), and the topological nature of the skyrmion lattice (represented by the $g$-term) is smeared out.
Any interaction Hamiltonian containing the magnetization to the second order does not give rise to such interaction connecting the variables for the three helices in the slowly-varying case.
These results are consistent with the observation by M\"uhlbauer et al. \cite{Muhlbauer09}.
On the other hand, skyrmion lattice is realized in numerical simulations using the Landau-Lifshitz-Gilbert equation \cite{Mochizuki12,Ohe13}, and it appears that skyrmion lattice arises without the ``cubic'' term.
We speculate that the relaxation process used to obtain thermalized state numerically is essential for the appearance of the skyrmion phase.
In fact, thermalization corresponds to the imaginary-time development, which is equivalent to taking account of the Ginzburg-Landau free energy, and thus the cubic term may be effectively included in the simulation.

\section{Effective Lagrangian and equations of motion}

As we have noted, $\beta_\pm$ in the Lagrangian (\ref{Lphibetascha}) play roles of canonical momenta for $\varphi_\pm$. 
By integrating out the two $\beta_\pm$ modes, we can write the Lagrangian  in terms of $\varphi_\pm$ only.
Integrals are easily calculated resulting in a contribution to the Lagrangian (see Appendix \ref{SECbetaint} for derivation)
\begin{align}
 \delta L_\varphi =\sumtd \frac{m_\varphi}{2}(\dot{\varphi}_+^2 +3 \dot{\varphi}_-^2),  \label{betaintL}
\end{align}
where 
\begin{align}
m_\varphi\equiv \frac{3\hbar^2}{2Jk^2},
\end{align}
is the mass for $\dot{\varphi}_+$.
The Lagrangian written in terms of $\varphi_\pm$ is thus 
\begin{align}
L &=  \sumtd \lt[2{\hbar}\Msk \overline{g} ( \varphi_+\dot{\varphi}_- - \varphi_- \dot{\varphi}_+) +\frac{m_\varphi}{2}(\dot{\varphi}_+^2 +3 \dot{\varphi}_-^2)
 -\frac{J\Msk^2}{3} \lt( (\nabla\varphi_+)^2+3(\nabla\varphi_-)^2 \rt)  
   \rt], \label{Lphi2}
\end{align}
The Lagrangian (\ref{Lphi2}) indicates that 
we have two modes with different mass and exchange constant and coupled by a kinetic term first order in time-derivative.
The factor of three difference is due to the difference of the magnitudes of the propagation vectors, $\kva+\kvb-2\kvc=-3\kvc$ and $\kva-\kvb=\sqrt{3} \hat{\zv}\times\kvc$.
In the Fourier representation, Eq. (\ref{Lphi2}) is
\begin{eqnarray}
L=\sumom\sumqv (\varphi_+,\varphi_-)_{-\qv,-\omega} 
\lt(\begin{array}{cc}
    \frac{J\Msk^2 q^2}{3}-\frac{m_\varphi \omega ^2}{2} & -2 i\hbar \Msk \overline{g} \omega \\
     2 i\hbar \Msk \overline{g} \omega & 3 \left(\frac{J\Msk^2 q^2}{3}-\frac{m_\varphi \omega
^2}{2}\right) \end{array}
              \rt)
\lt( \begin{array}{c} \varphi_+ \\ \varphi_-\end{array} \rt)_{\qv,\omega} 
\label{matrixphi2}
\end{eqnarray}
The energy dispersion determined from the determinant is 
\begin{align}
  \hbar \omega_q^\pm &= \frac{2\sqrt{2}\Msk \hbar^2 \geff}{\sqrt{3} m_\varphi}
  \lt[ 
    1+\frac{m_\varphi Jq^2}{4(\hbar \geff)^2} \pm \sqrt{  1+\frac{m_\varphi Jq^2}{2(\hbar \geff)^2}  } \rt]^{\frac{1}{2}}. \label{dispersionphi}
\end{align}
We thus obtain 
\begin{align}
  \hbar \omega_q^\pm &=  \lt\{ 
  \begin{array}{c} \frac{J\Msk}{2\sqrt{3}\geff }q^2 +O(q^4) \\ 
      \frac{8\geff J\Msk k^2}{3\sqrt{3}}+ \frac{J\Msk}{2\sqrt{3}\geff}q^2 .
  \end{array} \rt.
\end{align}
We see that the prefactor of the $q^2$ term of the massive mode is different from the one obtained in Eq. (\ref{dispersion_q2}). 
The discrepancy is due to the fact that we have neglected the contribution containing $q^2\partial_t^2$ in deriving Eq. (\ref{betaintL}).
We shall show in Appendix \ref{SECbetaint} that the correct dispersion is reproduced even after the integration  if we include the $q^2\partial_t^2$ contribution in $\delta L_\varphi$.
The effective Lagrangian of Eq. (\ref{Lphi2}) is of the form commonly used to discuss the low energy excitations, but it does not always describe the correct dispersion to the order of $q^2$.

Let us derive the equations of motion for $\varphi_\pm$.
In discussing the magnetization dynamics, dissipation of angular momentum, described by the Gilbert damping parameter $\alphaG$, is essential \cite{Chikazumi97}.
The Gilbert damping cannot be expressed in terms of a Lagrangian, but is represented by dissipation function \cite{TKS_PR08},
\begin{align}
  W\equiv  \frac{\hbar\alphaG}{2} \sumtd \dot{\Mv}^2,
\end{align}
and considering the equation of motion defined as 
$\displaystyle \frac{d}{dt}\frac{\delta L}{\delta \dot{q}_i}-\frac{\delta L}{\delta q_i}=-\frac{\delta W}{\delta \dot{q}_i}$, where $q_i$ represent generalized coordinates.
Dissipation function for the skyrmion lattice reads
\begin{align}
  W = \hbar\alphaG \sumtd \lt[  \frac{1}{2} [(\dot{\beta}_+)^2 + 3(\dot{\beta}_-)^2 ]
    + \frac{1}{3} [(\dot{\varphi}_+)^2 + 3(\dot{\varphi}_-)^2 ]
    + \frac{1}{6} (\dot{\varphi}_0)^2 \rt].
    \label{dissipationfunc}
\end{align}
The equation of motion including the Gilbert damping is
\begin{align}
  -2\hbar \Msk  \geff  \dot{\varphi}_+ -3m_\varphi \ddot{\varphi}_- -2\hbar\alphaG \dot{\varphi}_-
  +2J\Msk^2\nabla^2\varphi_-  &=0 \nnr
  2\hbar \Msk \geff\dot{\varphi}_- -m_\varphi \ddot{\varphi}_+  -\frac{2}{3}\hbar\alphaG \dot{\varphi}_+
  +\frac{2}{3}J\Msk^2\nabla^2\varphi_+  &=0 .
  \label{eqofmo0}
\end{align}


\section{Microwave response}

Let us discuss the microwave response of skyrmion lattice based on our low energy Lagrangian.
We consider two cases with the AC magnetic field, $B_{\rm ac}$, with the angular frequency of $\omega$ applied parallel and perpendicular to the $xy$ plane, respectively.
The applied magnetic field, $\Bv_{\rm ac}$,  couples to the magnetization as 
\begin{align}
  H_{B} = -\frac{g\mub}{a^3}\sumtd \Bv_{\rm ac}\cdot\Mv,
\end{align}
where the magnetization is expressed in terms of collective coordinates in Eqs. (\ref{skxconfiguration})(\ref{Mvi}).

\subsection{In-plane field ($B_{\rm ac}\perp \zv$)}
An in-plane AC field, $B_{\rm ac}^{\parallel}$, couples to the excitation as 
\begin{align}
  H_{B} = -\frac{g\mub}{2a^3}\sumtd B_{\rm ac}^{\parallel}
   \lt[ (\beta_++3\beta_-)\cos\phi_B+  \sqrt{3}(\beta_+-\beta_-)\sin\phi_B \rt],
\end{align}
where $\phi_B$ is the angle representing the field direction in the $xy$-plane.
Note that the in-plane field does not directly excite $\varphi_\pm$ modes because of rapidly oscillating components of magnetization.

A response to a microwave is described by the correlation function, given by the inverse of the Lagrangian (\ref{Lphibetascha}). 
The diagonal component of the correlation function for $\beta_\pm$, $\chi_{\beta_\pm}$ reads (including an imaginary part arising from the spin damping by $\hbar\omega\rightarrow\hbar\omega+i\eta$) 
\begin{align}
  \chi_{\beta_+} &=\frac{\frac{12}{81}
  \lt[3J\Msk q^2[(\hbar\omega)^2-(\hbar\omega_q^{\rm L})^2]+16(\hbar\omega)^2 J\Msk k^2\geff^2\muq\rt]}
  {[(\hbar\omega+i\eta)^2-(\hbar\omega_q^{+})^2] [(\hbar\omega+i\eta)^2-(\hbar\omega_q^{-})^2] },
\end{align}
and $ \chi_{\beta_-}=\frac{1}{3} \chi_{\beta_+}$, where $\omega_q^{\pm}$ are given by Eq. (\ref{dispersion}) with $g$ replaced by $\geff$.
The modes $\varphi_\pm$ are canonical conjugate of $\beta_\pm$ and thus the correlation functions for  $\varphi_\pm$ have the same poles. 
For $q=0$, the response to the in-plane AC field thus has a peak at $\omega=\omega_{q=0}^{+}=J\Msk k^2f=\Msk f \frac{D^2}{J}$ as well as at $\omega=0$.
For $J/a^2=1 $ meV and $D/a=0.18$ meV \cite{Iwasaki13} the resonant frequency is  $\nu\equiv \frac{\omega}{2\pi}=1.2$ GHz if $\Msk=1$ and $f=1$. 
The peak shifts to lower frequency when renormalization becomes larger (smaller $f$) by reducing the perpendicular magnetic field.  
The excitation mode described by $\beta_\pm$ corresponds to a shift of the skyrmion core breaking the rotationally symmetric skyrmion structure, and it can be thus identified with the rotational mode. 
Our analysis indicates that there is only one excitation energy for the rotation.
In contrast, in numerical simulation by Mochizuki \cite{Mochizuki12}, clockwise and counterclockwise 
two rotational modes having different frequencies were found.
The origin of discrepancy is not clear at present, but the boundary effect in the simulation may affect.

Experimentally, Onose et al. \cite{Onose12} observed an excitation of skyrmion lattice in Cu$_2$OSeO$_3$ by microwave absorption, and found an absorption peak around 1 GHz when the AC magnetic field is within the plane of the skyrmion lattice. It was also found that the peak shifts to higher frequency when the external field in the perpendicular direction increases.
Those observed behaviors seem consistent with the excitation of the present collective coordinate scenario, although our analysis is in two-dimensions while the experiment is in three-dimensional systems.

\subsection{$B_{\rm ac}\parallel \zv$}
When AC field is applied along the $z$ axis,
our Lagrangian, Eq. (\ref{Lphibeta2matrix}), indicates that the mode excited are $\varphi_0$ and $\mf$, the uniform component of the magnetization.
As we have seen, the mode $\varphi_0$ has no dynamic term in the Lagrangian, and thus its excitation occurs only at $\omega=0$. 
In contrast, the fluctuation of $\mf$ results in a peak at finite frequency.
The mode $\mf$ corresponds to an expansion and contraction of the core, and is identified with breathing mode pointed out by Mochizuki \cite{Mochizuki12}.
The existence of a single excitation mode  for $B_{\rm ac}\parallel \zv$ is consistent with the experimental result of Onose et al. \cite{Onose12}.

To analyze the response of $\mf$ is beyond the scope of the present paper, since we have treated $\mf$ as given. It can be carried out by minimizing the mean-field energy (See Appendix \ref{SECGL}) and by treating $\mf$ as a dynamic variable.


\section{Pinning model}

We consider a case of pinning due to  random impurities which modifies the local anisotropy energy.
The interaction we consider is 
\begin{align}
  H_{{\rm pin(z)}} &= - \sumtd v_{\rm p} \sum_{\Rv_i} \delta(\rv-\Rv_i) (M_z)^2 ,\label{Hpinz0}
\end{align}
where $\Rv_i$ is the position of random impurities and $v_{\rm p}$ is the pinning strength.

We here try to derive the effective pinning potential for $\varphi_i$'s following Fukuyama and Lee \cite{Fukuyama78,FukuyamaJPSJ78}.
We consider the case of modification of local anisotropy energy and write the integration over $\rv$ as a integral over a finite domain with area $S_0$ and summation over the domains (labeled by $\Rv$)  as 
\begin{align}
  H_{{\rm pin(z)}} &\simeq  - v_{\rm p} \sum_{\Rv}  \sum_{\Rv_i} 
  \lt( \sum_{j=a,b,c}\cos^2(\kv_j\cdot\Rv_i+\varphi_j(\Rv_i)) \rt. \nnr 
  &\lt. + \sum_{j,k=a,b,c}2\cos(\kv_j\cdot\Rv_i+\varphi_j(\Rv_i))\cos(\kv_k\cdot\Rv_i+\varphi_k(\Rv_i)) \rt)\nnr
  &= - \frac{v_{\rm p}}{4} \sum_{\Rv}  \sum_{\Rv_i} 
  \lt( \sum_{j=a,b,c}e^{i(2\kv_j\cdot\Rv_i+2\varphi_j(\Rv_i))} 
  + \sum_{j,k=a,b,c}2e^{i((\kv_j+\kv_k)\cdot\Rv_i+\varphi_j(\Rv_i)+\varphi_k(\Rv_i))} +{\rm c.c.} \rt).
\end{align}
The area $S_0$ is defined from the length scale where variation of $\varphi_i$ is smaller than unity.
Then the summation over impurity positions, $\Rv_i$, is carried out in each domain noting the fact that $\varphi_i$'s are treated as constant in the domain and that the summation over impurity positions results in a random walk as 
\begin{align}
  \sum_{\Rv_i} e^{i(2\kv_j\cdot\Rv_i+2\varphi_j(\Rv_i))}  & \simeq 
  \sqrt{N_{\rm i}} e^{i (2\varphi_j-\xi_j)} \nnr
  \sum_{\Rv_i} e^{i((\kv_j+\kv_k)\cdot\Rv_i+\varphi_j(\Rv_i)+\varphi_k(\Rv_i))} 
  & \simeq 
  \sqrt{N_{\rm i}} e^{i ((\varphi_j+\varphi_k)-\xi_{jk})} ,
\end{align}
where $N_{\rm i}$ is the number of impurities in a domain and $\xi_j$ and $\xi_{jk}$ represent a phase arising from random walk \cite{FukuyamaJPSJ78}.
The random potential (\ref{Hpinz0}) thus results in 
\begin{align}
  H_{{\rm pin(z)}} &\simeq 
  - \frac{v_{\rm p}}{4} \frac{\sqrt{N_{\rm i}}}{S_0} \int{d^2\Rv} \sum_{j=a,b,c} 
  \lt[\cos(2\varphi_j-\xi_j)+\sum_{k=a,b,c} 2 \cos ((\varphi_j+\varphi_k)-\xi_{jk})\rt].
\end{align}
Since we have six random phase factors, $\xi_j$ and $\xi_{jk}$, we cannot derive explicit expressions for $\varphi_i$'s which minimize the pinning energy. This fact is in contrast to the case of charge density waves with simply a single random phase, and we cannot therefore proceed further in the same manner as in Ref. \cite{Fukuyama78,Lee79}.
Nevertheless, we may assume that the effective pinning potential of harmonic shape is a good approximation if one regards $\varphi_\pm$ as the deviations of the phase variables from the local equilibrium configuration determined by the random pinning potential.
We thus consider an effective pinning potential
\begin{align}
  H_{{\rm pin}} &=  \sumtd \lt(\frac{K_+}{2}\varphi_+ ^2 + \frac{K_-}{2}\varphi_- ^2 \rt),\label{Hpin0}
\end{align}
where $K_\pm$ represent the strength of pinning of $\varphi_\pm$, respectively, which are treated as phenomenological parameters.
We assume further that the periodicity with respect to $\varphi_\pm$ persists in the pinning potential, namely, we assume 
\begin{align}
  H_{{\rm pin}} &=   - \sumtd \lt(K_+ \cos\varphi_+ + K_-\cos\varphi_-\rt).\label{Hpin}
\end{align}
We expect from the symmetry argument that $K_+=\frac{1}{3}K_-\equiv K_{\rm p}$, and we consider this case below.
Note that variables $\varphi_\pm$ here represent modifications of phase variables with respect to the equilibrium configuration determined taking account of random pinning potential.
The Lagrangian for skyrmion lattice including the pinning effect  is
\begin{align}
  L &=  \sumtd \lt[  -2\hbar \Msk \geff (\dot{\varphi}_+ \varphi_- - \dot{\varphi}_- \varphi_+)
     + \frac{m_\varphi}{2}(\dot{\varphi}_+^2 + 3 \dot\varphi_-^2)
     - \frac{J\Msk^2}{3}[(\nabla\varphi_+)^2+3(\nabla\varphi_-)^2]  \rt. \nnr
     & \lt.
     +K_{\rm p}\lt( \cos\varphi_+ + 3\cos\varphi_-\rt)\rt].  \label{Lphipin}
\end{align}
The excitation energy of skyrmion is modified by the pinning.
Considering a small oscillation in Eq. (\ref{Lphipin}), we obtain the dispersion relation 
\begin{align}
  \omega_q^\pm =\frac{1}{\sqrt{3}m_\varphi}
    \lt[ \sqrt{ (\hbar \Msk \geff)^2+4m_\varphi\lt(K_{\rm p}+\frac{2}{3}J\Msk^2 q^2\rt) } \pm (\hbar \Msk \geff) \rt].
\end{align}
Both of the two excitations thus become massive when pinning is present,  
\begin{align}
  \omega_q^\pm = \frac{ \hbar \Msk \geff}{\sqrt{3}m_\varphi}
    \lt[ \sqrt{ 1+\frac{3m_\varphi K_{\rm p}}{ (\hbar \Msk \geff)^2 } }\pm 1 \rt]+O(q^2).
\end{align}
The phenomenological parameter of the pinning strength, $K_{\rm p}$, is thus accessible experimentally by observing the excitation energy by microwave absorption.


\section{Spin transfer effect in metals }
So far we have considered a general case applicable to both metals and insulators.
In this section, we consider the metallic case and discuss the dynamics induced by an applied electric current.
In ferromagnetic metals, spin textures are usually slowly-varying compared to the conduction electron 
wavelength and the $sd$ exchange interaction coupling the conduction electron spin and localized moment is strong. 
In this case, the conduction electron spin going through the magnetization texture is rotated to align  along the local magnetization direction. 
This spin rotation exerts a torque on the magnetization called the spin-transfer torque \cite{Berger86,Slonczewski96}, and this torque is the main driving force for the slowly-varying structures. 

The spin-transfer effect is represented by the Hamiltonian 
\begin{align}
 H_{\rm ST} &\equiv \hbar\Msk\tilde{P} \sumtd (\cos\theta-1) (\jv\cdot\nabla){\phi} ,
\end{align}
where $\tilde{P}\equiv\frac{Pa^2}{2e\Msk}$, $P$ is the spin polarization of the current, and $\jv$ represents the current density \cite{TKS_PR08}.
In the same manner as the spin Berry's phase term in Sec. \ref{SECsbp}, we obtain an equivalent form suitable for studying the collective dynamics as 
\begin{align}
 \delta H_{\rm ST} &= \hbar\Msk\tilde{P} \sumtd \nv\cdot((\jv\cdot\nabla){\nv}\times\delta\nv).
\end{align}
By use of
\begin{align}
  \nabla_\mu {\nv} &= \sum_i\lt[ (\kv_i)_\mu (\hat\kv_i\times\nv_i)+\nabla_\mu {\beta}_i \hat \kv_i \rt],
\end{align}
where $\mu$ denotes spatial direction, 
we obtain 
\begin{align} 
 \delta H_{\rm ST} &= 
   \hbar\Msk\tilde{P}g \sumtd 
\lt[
((\jv\cdot\kv_a) \varphi_b- (\jv\cdot\kv_b)\varphi_a )
+( (\jv\cdot\kv_b)\varphi_c- (\jv\cdot\kv_c) \varphi_b )
+( (\jv\cdot\kv_c) \varphi_a- (\jv\cdot\kv_a) \varphi_c )\rt] \nnr
& \times
\cos (\varphi_a+\varphi_b+\varphi_c) 
\nnr
&+ 
\tilde{P} \sumtd
\lt[
(\jv\cdot\nabla{\beta}_a) \lt(\varphi_a- \frac{1}{2}\lt(\varphi_b + \varphi_c\rt)\rt)
+
(\jv\cdot\nabla{\beta}_b) \lt(\varphi_b- \frac{1}{2}\lt(\varphi_c + \varphi_a\rt)\rt)
\rt. \nnr
& \lt. +
(\jv\cdot\nabla{\beta}_c) \lt(\varphi_c- \frac{1}{2}\lt(\varphi_a + \varphi_b\rt)\rt)
\rt].
\end{align}
The dominant terms are those including only  $\varphi_i$'s, since $\beta$-modes have mass gap.
Using the self-consistent harmonic approximation and in terms of $\varphi_\pm$, they read 
\begin{align} 
 \delta H_{\rm ST} &=    \hbar\Msk\tilde{P} \sumtd  
2\geff(v_-\varphi_+ - v_+\varphi_- ),
\end{align}
where 
\begin{align} 
 v_- & \equiv \frac{\tilde{P} }{2}\jv\cdot(\kva-\kvb)=\tilde{P} \frac{\sqrt{3}}{2}\jv\cdot(\hat{\zv}\times\kvc) \nnr
 v_+ & \equiv  \frac{\tilde{P} }{2} \jv\cdot(\kva+\kvb-2\kvc)=-\tilde{P} \frac{3}{2} \jv\cdot\kvc ,
\end{align}
are driving speed of $\varphi_\pm$.
We see that the spin-transfer effect is renormalized by the fluctuation of $\varphi_0$ by a factor of $f$.

The total Lagrangian for a skyrmion lattice in low energy region in metals including the pinning and the spin-transfer effect  is
\begin{align}
  L &=  \sumtd \lt[  -2\hbar \Msk \geff (\dot{\varphi}_+ \varphi_- - \dot{\varphi}_- \varphi_+)
     + \frac{m_\varphi}{2}(\dot{\varphi}_+^2 + 3 \dot\varphi_-^2)
     - \frac{J\Msk^2}{3}[(\nabla\varphi_+)^2+3(\nabla\varphi_-)^2]  \rt. \nnr
     & \lt.
     -2 \hbar\Msk\geff (v_-\varphi_+ - v_+\varphi_- )
     +K_{\rm p}\lt( \cos\varphi_+ + 3\cos\varphi_-\rt)\rt].  \label{Lphitotal}
\end{align}
We see that the external current couples to the phase $\varphi_\pm$ itself. This might seem unphysical since the absolute value of phase has no physical meaning, but is correct, as is known in the case of domain walls \cite{TKS_PR08,TK06}. We shall indeed show below that the absolute value of phase does not appear in the equation of motion (see Eq. (\ref{eqofmo})).
This Lagrangian is essentially an extension of the single-particle Lagrangian for a single magnetic vortex \cite{SNTKO06} into the case of lattice structure if the two-dimensional coordinates $(X(t),Y(t))$ representing the core of vortex are replaced by two field variables, $\varphi_\pm(\rv,t)$.
In fact the single skyrmion dynamics is essentially the same as a single magnetic vortex represented by the Thiele equation. 
The dynamics of a magnetic domain wall is, in contrast, very different form skyrmions and vortices.
In fact, a significant feature of skyrmion lattices is that they can be driven at much lower current density than magnetic domain walls \cite{Schulz12}.
The Lagrangian for a domain wall is described by two coordinates, $X(t)$ representing the position, and $\phi(t)$ representing the angle of the wall plane and these variables play roles of $\varphi_\pm$ \cite{TK04}. The translational motion of the wall therefore requires the angle $\phi$ to grow, and the development of $\phi$ costs energy because of hard-axis magnetic anisotropy energy. 
This results in the intrinsic pinning effect which prevents the wall to move at low current. 
Low threshold current is thus realized either by introducing non-adiabatic torque \cite{Thiaville05,Li04st} or by lowering the hard-axis anisotropy \cite{Koyama11}.
In contrast, in the case of vortices and skyrmions, translation modes in the two spatial directions form canonical conjugates. The motion induced is therefore without energy cost if in the absence of pinning \cite{TKS_PR08}.
The lower current density for vortices and skyrmions is thus understood qualitatively based on the Lagrangian.

It has been  known that the current generally induces another torque on magnetization, called the non-adiabatic torque, which is perpendicular to the spin-transfer torque \cite{TKS_PR08}.
This torque arises from  spin relaxation and non-adiabatic scattering of conduction electrons and it cannot be expressed in terms of a Hamiltonian similarly to the case of the Gilbert damping.
The torque is represented in the equation of motion for spin by replacing the Gilbert damping term, $\alphaG\dot{\nv}$ by
$[\alphaG\partial_t-\betana\tilde{P}(\jv\cdot\nabla)]\nv$, where $\betana$ is a dimensionless parameter representing the non-adiabaticity and spin relaxation \cite{TK04,Zhang04,Thiaville05,TKS_PR08}.
In terms of phason, this corresponds to replacing 
 $\alphaG\dot{\varphi}_\pm$ by
$\alphaG\dot{\varphi}_\pm-\betana v_\pm$, since a spatial derivative gives rise to a factor proportional to the phason wave vector.

Considering the case where the current is applied for $\varphi_+$, i.e., $v_-=0$,  
the equation of motion including the Gilbert damping and non-adiabatic torque is therefore  
\begin{align}
  -2\hbar \Msk  \geff ( \dot{\varphi}_+ - v_+)-3m_\varphi \ddot{\varphi}_- -2\hbar\alphaG \dot{\varphi}_-
  +2J\Msk^2\nabla^2\varphi_- - 3K_{\rm p} \sin \varphi_- &=0 \nnr
  2\hbar \Msk \geff\dot{\varphi}_- -m_\varphi \ddot{\varphi}_+  
   -\frac{2\hbar}{3}\lt(\alphaG \dot{\varphi}_+ -\betana v_+\rt)
  +\frac{2}{3}J\Msk^2\nabla^2\varphi_+ - K_{\rm p} \sin \varphi_+ &=0 .
  \label{eqofmo}
\end{align}
From the equation of motion, Eq. (\ref{eqofmo}), the threshold current $\jc$ for the motion of the skyrmion lattice, determined by $2\hbar \Msk \geff v_+ =3K_{\rm p}$, is 
\begin{align}
  \jc= \frac{2eK_{\rm p}}{\hbar Pa^2 k fg},\label{jcresult}
\end{align}
where $e$ is the elementary electric charge.
The threshold current thus increases when the renormalization factor $f$ decreases. 

After depinning, there is a solution of a steady flow if mass term is neglected. In the case of $\nabla\varphi_\pm=0$, the solution is 
\begin{align}
  \dot\varphi_+&=v_+ \frac{1+\frac{\alphaG\betana}{3(\Msk\geff)^2}}{1+\frac{\alphaG^2}{3(\Msk\geff)^2}}\nnr
  \dot\varphi_-&=v_+ \frac{\alphaG-\betana }{3\Msk\geff}\frac{1}{1+\frac{\alphaG^2}{3(\Msk\geff)^2}}.
\end{align}
The longitudinal and transverse velocities of the lattice is defined as 
$v_\parallel= \dot\varphi_+/k$ and $v_\perp= \dot\varphi_-/k$, respectively.
we see that the transverse velocity is proportional to $1/f$ as a result of renormalization.
(The result in the absence of renormalization effect was obtained previously in Ref. \cite{IwasakiNC13}.)
Since both $\alphaG$ and $\betana$ are usually small (typically of the order of $10^{-2}$), the longitudinal terminal velocity appears to be insensitive to $\beta$, as noted in Ref. \cite{IwasakiNC13} and in Ref. \cite{Kasai06} in the case of magnetic vortex .

To reproduce the observed three-dimensional threshold current density, defined as $\jc/a$, of $10^6$ A/m$^2$ \cite{Schulz12}, Eq. (\ref{jcresult}) indicates a very weak collective pinning energy of  $K_{\rm p}\sim 6\times 10^{-11}$ eV if $a=5$\AA, $P=1$, $J/a^2=1$ meV, $D/a=0.18$ meV \cite{IwasakiNC13}.
It was shown  in bulk MnSi that the threshold current density increases rapidly as a function of the temperature near the critical temperature \cite{Schulz12}.
It was discussed there that the behavior is not explained solely by the descrease of $\Msk$, and that the enhancement of the pinning force due to softening of the skyrmion lattice needs to be taken into account.
The renormalization effect we have found, indicating that the effective pinning force, proportional to $K_{\rm p}/f$, increases at higher temperatures due to the factor of $1/f$ may explain partially the observed softening behavior.
The opposite temperature dependence was observed in a thin FeGe and the result was argued to be due to the weakening of the pinning by the thermal fluctuation \cite{Yu12}.

\section{Topological Hall effect}

The topological charge of magnetic structures induces the Lorentz force on the conduction electrons and induces the Hall effect, called the topological Hall effect. 
The Hall force in a film is expressed generally as  \cite{TKS_PR08}
\begin{align}
  F_{{\rm H},i} &=  \frac{\pi\hbar}{e}Pd \sum_j j_j \Phi_{ij},
\end{align}
where $i$ denotes a spatial direction, $-e(<0)$ is the electron charge, $P$ is the spin polarization of the current, $d$ is the thickness of the film and
\begin{align}
  \Phi_{ij}\equiv \frac{1}{4\pi M^3}\int d^2 r \Mv\cdot(\nabla_i \Mv\times\nabla_j\Mv),
\end{align}
is the topological charge defined in the $ij$-plane. 
The Hall resistivity is
\begin{align}
  \rho_{xy} =-\frac { F_{{\rm H},y} } {ej_x} =-\frac{\pi\hbar}{e^2}Pd\Phi_{xy}.
\end{align}
In the case of skyrmion lattice, described by Eq. (\ref{nvidef}), the topological charge is 
calculated as (see Eq. (\ref{SBPvalue}))
\begin{align}
\Phi_{xy}
&=g\cos \varphi_0 =\geff,
\end{align}
within the self-consistent harmonic approximation.
The magnitude of the topological Hall effect is renormalized by a factor of $f(=\geff/g)$ in the skyrmion lattice phase.

\section{Conclusion}
We have presented a theoretical description of a two-dimensional skyrmion lattice realized in helical magnets by use of phason fields.
The ground state skyrmion lattice is described in terms of three helices following the previous approaches by M\"uhlbauer et al.  and  Petrova et al. \cite{Muhlbauer09,Petrova11}, and we have introduced collective coordinates consisting of fluctuations of helix phases (phasons) and perpendicular fluctuations.
By deriving an effective Lagrangian describing slowly-varying phasons, 
we have confirmed previous observations that there are two excitation modes, one gapless mode having quadratic dispersion and a massive mode. 
The vector nature of spin does not lead to an essential difference from the scalar field like in charge-density waves (CDW) as for the excitation concerns except for the fact that spin phason is always coupled to an effective magnetic field of the spin Berry's phase. 
We have found that there is another phase variable, $\varphi_0$, governing the stability and the topological nature of the skyrmion lattice.
We demonstrated that the fluctuation of this mode results in a screening of the topological charge of skyrmion lattice, and that the screening effect would be observable in various measurements such as the microwave absorption, and current-induced dynamics and topological Hall effect in metals.

\begin{acknowledgment}
The authors  thank X. Z. Yu, S. Seki, J. Kishine, C. Marrows, M. Mochizuki, W. Koshibae, A. Beekman, D. Takahashi and M. Ogata for valuable comments and discussions. 
H. F. thanks N. Nagaosa for useful discussions in early stage and J. Kishine for drawing attention to Ref. \cite{Petrova11}.
This work was supported by a Grant-in-Aid for Scientific Research (C) (Grant No. 25400344) and (A) (Grant No. 24244053) from Japan Society for the Promotion of Science and UK-Japanese Collaboration on Current-Driven Domain Wall Dynamics from JST.
\end{acknowledgment}

\appendix
\section{Mean-field energy of skyrmion lattice
\label{SECGL}}
In this section we summarize the mean-field energy  without applied current and pinning estimated for the skyrmion lattice structure, represented by 
Eqs. (\ref{skxconfiguration})(\ref{nvi0def}).
The energy per unit site evaluated for the free energy (\ref{Hamiltonian}) without both pinning and current-induced torque is 
\begin{align}
  E_{\rm sk} &= \frac{a_M}{2}\mf^2+\frac{b_M}{4}\mf^4+\frac{2\mub}{a^3}B\mf 
   -\frac{\alpha_M}{2}\Msk^2+\frac{\beta_M}{4}\Msk^4 
   +\gamma_1 \mf\Msk^3 +\gamma_2 \mf^2\Msk^2, \label{GLenergy}
\end{align}
where $\alpha_M=3\lt(\frac{D^2}{J}-a_M\rt)$, $\beta_M=\frac{51}{4}b_M$, $\gamma_1=\frac{9}{4}b_M$ and $\gamma_2=3b_M$. The term $\gamma_1$ is essential to stabilize the skyrmion structure.

For comparison, the energy for a single helix is given also by Eq. (\ref{GLenergy}) but with different parameters; 
 $\alpha_M=\lt(\frac{D^2}{J}-a_M\rt)$, $\beta_M=\frac{1}{4}b_M$, $\gamma_1=0$ and $\gamma_2=b_M$.

\section{Integrating out degrees of freedom}
Here we describe briefly the integrating-out variables by carrying out path-integral over variables. Usually this is carried out to focus on variables describing energy dynamics, and it is also useful to switch between $p,q$-representation to $q,\dot{q}$-representation. 
It should be noted that the calculation keeps all the quantum fluctuations included in the result if done without approximation. It is not thus equivalent to deleting variables in the equations of motion. 

We first consider a case of a particle in one-dimension, whose  Hamiltonian is $H=\frac{p^2}{2m}+V(q)$, where $p$ is the canonical momentum for $q$, $m$ is the mass of the particle and $V$ is a potential.  The equations of motion (Hamilton equations) read 
\begin{align}
  \dot{p} &= - \frac{\delta H}{\delta q} = - \frac{d V(q)}{d q} \nnr
  \dot{q} &= \frac{\delta H}{\delta p} = \frac{p}{m} . \label{hamiltoneq}
\end{align}
The equations are also obtained from the Lagrangian, defined as $L(q,p)=p \dot{q}-H$.
In the path-integral formalism \cite{Feynman65}, the dynamics of $q$ and $p$ is represented by a functional integral (denoted by ${\cal D}$) called the partition function, defined as
\begin{align}
  Z & \equiv \int {\cal D}q {\cal D}p e^{i\int dt (p \dot{q}-H)}.
\end{align}
In the present example, the Lagrangian is quadratic with respect to $p$ and thus the integration over $p$ results in  
\begin{align}
  Z & = \int {\cal D}q e^{i\int dt L(\dot{q},q))},
\end{align}
where $L(\dot{q},q)\equiv \frac{m}{2}\dot{q}^2-V(q)$ and we have dropped an irrelevant constant.
We thus obtained the Lagrangian represented by $\dot{q}$ and $q$. 
Since the integral was carried out exactly, the resulting Lagrangian is quantum mechanically equivalent to the original Hamiltonian. At the classical level, the equation of motion obtained from 
$ L(\dot{q},q))$, $ m\ddot{q} = - \frac{d V(q)}{d q}$, is equivalent to Eq. (\ref{hamiltoneq}).

\subsection{Integrating-out $\varphi_0$ \label{SECphiint}}

The partition function for $\varphi_0$ is written in the imaginary-time path integral formalism as \cite{Sakita85} 
\begin{align}
  Z_{\varphi_0} &= \int{\cal D}\varphi_0 \! e^{\beta L_{\varphi_0}},
\end{align}
where we noted that the Hamiltonian for $\varphi_0$ is $-L_{\varphi_0}$ and  $\beta\equiv (\kb T)^{-1}$.
The expectation value $\overline{{\varphi_0}^2}$, defined as
\begin{align}
 \overline{{\varphi_0}^2}\equiv \frac{1}{ Z_{\varphi_0}} 
    \int{\cal D}\varphi_0  \varphi_0^2(\rv) e^{-\beta L_{\varphi_0}},
\end{align}
reads by using Fourier transform, 
\begin{align}
 \overline{{\varphi_0}^2} 
 &=\frac{1}{ Z_{\varphi_0}} 
    \int{\cal D}\varphi_0 \sumqv |\varphi_0(\qv)|^2 
       e^{ -\beta\sum_{\qv'} \frac{1}{2}\lt[\frac{J}{3}(q')^2+hf\rt]|\varphi_0(\qv')|^2  } \nnr
       &=
   - \frac{\delta  \ln Z_{\varphi_0}} {\delta \lt[ \frac{\beta}{2}\lt(\frac{J}{3}q^2+hf\rt)\rt]}.
\end{align}
The partition function is a Gaussian integral, and thus 
\begin{align}
  Z_{\varphi_0} &= \prod_{\qv} \pi \lt[\frac{\beta}{2}\lt(\frac{J}{3}q^2+hf\rt)\rt]^{-1},
\end{align}
and thus we obtain 
\begin{align}
 \overline{{\varphi_0}^2} 
 &= 2\kb T \sumqv \frac{1}{\frac{J}{3}q^2+hf}.
\end{align}

\subsection{Integrating-out $\beta_\pm$ \label{SECbetaint}}
In the Lagrangian (\ref{Lphibetascha}), the contributions including $\beta_\pm$ are 
\begin{align}
L_\beta & \equiv \Msk^2 \sumtd \lt[-\frac{\hbar}{\Msk}( \beta_+ \dot{\varphi}_+ + 3 \beta_- \dot{\varphi}_- )
 - \frac{J}{3} \lt( k^2 ( \beta_+^2+3\beta_-^2) \rt)
                     -\frac{J}{2} [(\nabla\beta_+)^2+3(\nabla\beta_-)^2 ] \rt] .
\end{align}
The partition function in the real-time is defined as 
\begin{align}
  Z_\beta\equiv \int{\cal D}\beta_+ {\cal D}\beta_- e^{\frac{i}{\hbar}\int dt L_\beta}.
\end{align}
Using Fourier transformation, we carry out the integral as 
\begin{align}
  Z_\beta &= \int{\cal D}\beta_+ {\cal D}\beta_- 
  {\rm exp}\lt[-\frac{i}{\hbar}\int dt \sumqv \lt[\frac{J\Msk^2 k^2}{3} \lt(1+\frac{3q^2}{2k^2}\rt)  
  \rt.\rt.\nnr
  & \times 
  \lt(\lt|\beta_+(\qv)-\frac{3\hbar\dot{\varphi}_+(\qv)}{2J\Msk k^2\lt(1+\frac{3q^2}{2k^2}\rt)}\rt|^2 +3 \lt|\beta_-(\qv)-\frac{3\hbar\dot{\varphi}_-(\qv)}{2J\Msk k^2\lt(1+\frac{3q^2}{2k^2}\rt)}\rt|^2\rt) 
 \nnr
 & \lt. \lt.
  -\frac{3\hbar^2}{4Jk^2\lt(1+\frac{3q^2}{2k^2}\rt)}(\dot{\varphi}_+^2 +3 \dot{\varphi}_-^2)\rt] \rt]\nnr
  &\equiv e^{\frac{i}{\hbar}\int dt \delta L_\varphi}   ,  \label{betaintresult}
\end{align}
where $\delta L_\varphi$ is the effective Lagrangian arising from the $\beta$-integral.
The integral over $\beta_+$ is carried out as 
\begin{align}
  \int{\cal D}\beta_+ 
  {\rm exp}\lt[-\frac{i}{\hbar}\int dt \sumqv \frac{J\Msk^2 k^2}{3} \lt(1+\frac{3q^2}{2k^2}\rt)  
  \lt|\beta_+(\qv)-\frac{3\hbar\dot{\varphi}_+(\qv)}{2J\Msk k^2\lt(1+\frac{3q^2}{2k^2}\rt)}\rt|^2 \rt] \nnr
  = \prod_{\qv,t} \lt(\frac{3\pi\hbar}{i J\Msk^2 (k^2+3q^2)}\rt) ,  
\end{align}
which is a constant independent on dynamic variables. 
Neglecting irrelevant constants, the Lagrangian is  
\begin{align}
\delta L_\varphi
  = \sumqv \frac{m_\varphi}{2}\frac{1}{1+\frac{3q^2}{2k^2}} (\dot{\varphi}_+(\qv)^2 +3 \dot{\varphi}_-(\qv)^2),  \label{betaintL0}
\end{align}
where 
$m_\varphi\equiv \frac{3\hbar^2}{2Jk^2}$ is the mass for $\dot{\varphi}_+$.
In the real-space representation, the Lagrangian is non-local, but for discussing low energy dynamics of $\varphi_\pm$, it is enough to neglect the order of $q^2$, resulting in Eq. (\ref{betaintL}).

To discuss the massive excitation mode, the order of $q^2$ in Eq. (\ref{betaintL0}) needs to be kept.
In this case, we have 
\begin{align}
\delta L_\varphi
  = \sumtd \lt[   \frac{m_\varphi}{2} 
   (\dot{\varphi}_+(\qv)^2 +3 \dot{\varphi}_-(\qv)^2) 
    -  \frac{3m_\varphi}{4k^2} ((\nabla\dot{\varphi}_+)^2 +3 (\nabla\dot{\varphi}_-)^2) \rt] 
    +O(q^4\omega^2).  \label{betaintLexp}
\end{align}
Although the second term containing $\nabla^2\partial_t^2$ is higher-order contribution when considering standard excitations, it modifies the dispersion relation in the present case.
In fact, the matrix of Eq. (\ref{dispersionphi}) now reads
\begin{eqnarray}
\lt(\begin{array}{cc}
    \frac{J\Msk^2 q^2}{3}-\frac{m_\varphi \omega ^2}{2\mu_q} & -2 i\hbar \Msk \geff \omega \\
     2 i\hbar \Msk \geff \omega & 3 \left(\frac{J\Msk^2 q^2}{3}-\frac{m_\varphi \omega
^2}{2\mu_q}\right) \end{array}
              \rt),
\end{eqnarray}
where $\mu_q\equiv 1+\frac{3q^2}{2k^2}$.
The determinant of the matrix is (dropping constants)
\begin{align}
  \lt({\hbar\omega}\rt)^4
  -\lt({\hbar\omega}\rt)^2 \frac{64}{27}(J\Msk k^2\geff\mu_q)^2\lt(1+\frac{3q^2}{8k^2\geff^2\mu_q}\rt)
  +\frac{16}{81}(J^2\Msk^2 k^2q^2\mu_q)^2.\label{determinant2}
\end{align}
(This result is identical to Eq. (\ref{determinant_1}) if $g$ is replaced by $\geff=fg$.)
The energy dispersion determined from Eq. (\ref{determinant2}) is 
\begin{align}
  \hbar \omega_q^\pm &= \frac{4\sqrt{2}\Msk Jk^2 \geff\mu_q}{3\sqrt{3}}
  \lt[ 
    1+\frac{3q^2}{8(k\geff)^2\mu_q} \pm \sqrt{  1+\frac{3q^2}{4(k\geff)^2\mu_q}  } \rt]^{\frac{1}{2}}.
\end{align}
We thus obtain the massive mode as 
\begin{align}
  \hbar \omega_q^\pm &=  \frac{8\Msk Jk^2 \geff\mu_q}{3\sqrt{3}}
  \lt[ 
    1+\frac{3q^2}{8(k\geff)^2\mu_q}  \rt]
     \nnr
 &=\frac{8 \Msk Jk^2 \geff\mu_q}{3\sqrt{3}} +\frac{J\Msk(1+8\geff^2) }{2\sqrt{3} \geff} q^2 +O(q^4).
\end{align}
We therefore reproduce the result of Eq. (\ref{dispersion}) obtained before the integration over $\beta_\pm$.


\end{document}